\documentclass{aa}  

\usepackage{graphicx}
\usepackage{txfonts}
\usepackage[T1]{fontenc}
\usepackage{ae,aecompl}

\usepackage{amsmath}
\usepackage{amssymb}
\usepackage{multirow,bigdelim}
\usepackage{mathtools}
\usepackage{siunitx}
\usepackage[colorlinks=true,linkcolor=blue, citecolor=blue]{hyperref}%

\usepackage{subfig}

\newcommand{\mpch}{\,h^{-1}{\rm{Mpc}}}
\newcommand{\rp}{r_{\rm p}}
\newcommand{\zl}{z_{\rm l}}
\newcommand{\zs}{z_{\rm s}}
\newcommand{\dd}{\mathrm{d}}
\newcommand{\esd}{\Delta \Sigma}

\newcommand{\lum}{\texttt{luminous}}
\newcommand{\dense}{\texttt{dense}}
\newcommand{\lensfit}{\textit{lens}fit}

\defcitealias{Fortuna et al. 2021}{F21}

\usepackage{bm}
\usepackage{soul}
\usepackage{comment}

\usepackage{color}

\numberwithin{equation}{section}

\usepackage{etoolbox}

\begin{document}

\title{KiDS-1000: Weak lensing and intrinsic alignment around luminous red galaxies
}

\titlerunning{KiDS-1000 GGL of the LRG sample}

\author{
Maria Cristina Fortuna\inst{1},
Andrej Dvornik\inst{2}\thanks{\emph{E-mail:} dvornik@astro.rub.de},
Henk Hoekstra\inst{1},
Nora Elisa Chisari\inst{3,1},
Marika Asgari\inst{4},
Maciej Bilicki\inst{5},
Catherine Heymans\inst{6,2},
Hendrik Hildebrandt\inst{2},
Koen Kuijken\inst{1},
Angus H. Wright\inst{2},
Ji Yao\inst{7}
 }

\authorrunning{M.C. Fortuna et al. }

\institute{Leiden Observatory, Leiden University, PO Box 9513, Leiden, NL-2300 RA, the Netherlands \and
Ruhr University Bochum, Faculty of Physics and Astronomy, Astronomical Institute (AIRUB), German Centre for Cosmological Lensing, 44780 Bochum, Germany \and
Institute for Theoretical Physics, Utrecht University, Princetonplein 5, 3584CC Utrecht, The Netherlands \and
School of Mathematics, Statistics and Physics, Newcastle University, Herschel Building, NE1 7RU, Newcastle-upon-Tyne, UK \and
Center for Theoretical Physics, Polish Academy of Sciences, al. Lotników 32/46, 02-668 Warsaw, Poland \and
Institute for Astronomy, University of Edinburgh, Royal Observatory, Blackford Hill, Edinburgh, EH9 3HJ, UK \and
Shanghai Astronomical Observatory (SHAO), Nandan Road 80, Shanghai 200030, China 
}

\date{Accepted XXX. Received YYY; in original form ZZZ}

\label{firstpage}
\makeatletter
\renewcommand*\aa@pageof{, page \thepage{} of \pageref*{LastPage}}
\makeatother

\abstract{We study the properties of luminous red galaxies (LRGs) selected from the fourth data release of the Kilo Degree Survey (KiDS-1000) via galaxy-galaxy lensing of the background galaxies from KiDS-1000. We used a halo model formalism to interpret our measurements and obtain estimates of the halo masses as well as the satellite fractions of the LRGs, resulting in halo masses of  $2.7 \times 10^{12} h^{-1} {\rm M}_{\odot}<M_{\rm h}< 2.6 \times 10^{13} h^{-1} {\rm M}_{\odot}$. We studied the strength of intrinsic alignments (IA) using the position-shape correlations as a function of LRG luminosity, where we used a double power law to describe the relation between luminosity and halo mass to allow for a comparison with previous works. Here, we directly linked the observed IA of the (central) galaxy to  the mass of the hosting halo, which is expected to be a fundamental quantity in establishing the alignment. We find that the dependence of the IA amplitude on halo mass is described well by a single power law, with an amplitude of $A = 5.74\pm{0.32}$ and slope of $\beta_M = 0.44\pm{0.04}$, in the range of $1.9 \times 10^{12}h^{-1} {\rm M}_{\odot}<M_{\rm h}<3.7 \times 10^{14} h^{-1} {\rm M}_{\odot}$. We also find that both red and blue galaxies from the source sample associated with the LRGs are  randomly oriented, with respect to the LRGs, although our detection significance is limited by the uncertainty in our photometric redshifts.}


\maketitle



\section{Introduction}

The intrinsic alignment (IA) of galaxies, defined as the tendency of galaxies to point in a coherent direction, has gained attention in the last two decades as an important contaminant in lensing \citep[][and references therein]{Troxel2015review}. 
N-body simulations have explored the origin of the triaxility and angular momentum of dark matter halos and how they orient their major axis in the direction of matter over-densities, finding that the orientation also depends on the environment  and the location on the large-scale structure \citep[][]{Dubinski1992, Croft&Metzer2000, Hopkins2005, Oliver&Porciani2007,Lee2008, Forero-Romero2014}. Models of galaxy alignment predict that the galaxy inherits the orientation of its major axis from the orientation of the parent halo: such a relation is expected to be primarily sourced by the effect of tidal fields during galaxy formation \citep[][]{Catelan2001}. These models predict a dichotomy in the alignment of elliptical, pressure-supported galaxies and disc-like, rotationally supported galaxies \citep[e.g.][]{Catelan2001, Hirata2004, Blazek2017, Vlah2020}. This is in agreement with the results from observations \citep{Hirata2007, Faltenbacher2009, Joachimi2011b, Blazek2011, Singh2015, Johnston2019, Mandelbaum2011WiggleZ}.  

More specifically, observations have shown that luminous red galaxies (LGRs), as a sample of elliptical, pressure-supported galaxies, show strong IA signals and that the dependence on luminosity can be described by a power law \citep[][]{Hirata2007, Joachimi2011b, Singh2015}.  \citet{Fortuna2021b} extended observational constraints on the IA amplitude to samples of LRGs with significantly lower luminosities, compared to previous studies, and found that the dependence with luminosity is shallower, compared to the high-$L$ samples. This points towards a more complex behaviour in the luminosity-alignment relation.

The relation between luminosity and halo mass itself is complex. Hence, a simple dependence with halo mass, as predicted by models of IA \citep[][]{Xia2017, Piras2018}, would result in a complex dependence of the IA signal with luminosity. It is therefore interesting to explore the observational link between the IA of galaxies and the mass of the hosting halo. This could simplify the modelling and help generate synthetic galaxy catalogues that reproduce the observed IA using halo catalogues from N-body simulations \citep[e.g.][]{Carretero2017CosmoHub}. A direct measure of IA as a function of halo mass avoids the intermediate step of calibrating IA on secondary galaxy properties that are not directly responsible for the alignment mechanism, have non-negligible scatter, and may introduce a bias.

In this paper, we determine the average halo mass for the samples used in \citet[][]{Fortuna2021b}, directly linking their IA signal and their halo mass. \citet[][]{Singh2015, vanUitert2017} and \citet[][]{Piras2018} have addressed the same question using higher luminosity NS higher mass samples, finding a single power law relation for the IA dependence on halo mass. Here, we extend the analysis to the faint-end, where the IA dependence on luminosity changes its slope. This allows us to address the question of whether the observed flattening in the IA amplitude is a consequence of a similar flattening in the halo mass-luminosity relation.

To measure the halo masses, we employ weak gravitational lensing: light bundles of the distant galaxies are deflected by the matter distribution along the line of sight, which leads to an apparent correlation between the shape of a background galaxy (source) and the position of a foreground galaxy (lens). This galaxy-galaxy lensing (GGL) signal is an important tool to investigate the dark matter distribution around galaxies \citep[e.g.][]{Hoekstra2005,  Mandelbaum2006a}. 

In this paper, we use the halo occupation distributions based halo model to connect the statistical properties of dark matter halos to those of the galaxies. It is an analytical approach to predict observable quantities based on the link between the galaxy occupation statistics, the abundance, and the clustering of dark matter \citep[][]{Seljak2000, MaFry2000, CooraySheth2002, vandenBosch2013-PaperI, 2023Asgari_HaloModelReview}. 

In this study, the lens sample consists of LRGs in the footprint of the fourth data release of KiDS \citep[KiDS-1000,][]{Kuijken2019DR4}, which were selected via a variation of the \textsc{redMagiC} algorithm \citep[][]{Rozo2016redMaGiC}, as presented in \citet[][]{Vakili2023}. To measure the lensing signal, we used the source sample introduced in \citet[][]{Giblin2021}.

Luminous red galaxies are typically the central galaxies residing in massive halos and they are responsible for the alignment at large scales. The alignment at small scales is caused by satellite galaxies: the intra-halo tidal fields align the satellites by a torquing mechanism that leads to a net radial alignment signal towards the halo centre \citep[][]{Pereira2008}. In the second part of this paper, we investigate this signal by determining the alignment of galaxies in the source sample. These are selected to be physically close to the LRGs, so that the LRG can be considered a proxy for the halo centre. 
To do so, we employed an estimator that is similar to the one used for GGL, following the approach presented in \citet[][]{Blazek2012}. When measuring the signal, we accounted for the lensing contamination and the lensing dilution occurring between galaxies that are physically associated. The IA of the sources presented here is complementary to the study of the IA signal in \citet{Fortuna2021b}, who looked at the alignment signal of the LRG sample at large scales. Here, we constrained the small scale signal ($\rp<10 h^{-1} \rm Mpc$) sourced by non-LRG galaxies. By studying the mass dependence of the IA signal at large scales and its strength at small scales through the analysis of the alignment of the source galaxies surrounding the LRGs, we provide insights that can be used to model the IA signal in lensing studies. In particular, these results can be used to identify the dominant terms in the IA halo model \citep{SchneiderBridle2010, fortuna2020halo}.

The paper is organised as follows. In Sect.~\ref{sec:data}, we present the data employed in this work. In Sect.~\ref{sec:galaxy_galaxy_lensing}, we introduce the estimator used to measure the signal, while in Sect.~\ref{sec:model}, we present the model framework we use to interpret the measurements. In Sect.~\ref{sec:fitting_procedure}, we detail the fitting procedure and in Sects.~\ref{sec:lensing_results} and Sect.~\ref{sec:IA_results}, we present our results. In Sect. \ref{sec:conclusions}, we draw our conclusions.
Throughout the paper, we assume a flat $\Lambda$CDM cosmology with $h=0.7, \Omega_{\rm m} = 0.25, \Omega_{\rm b} = 0.044, \sigma_8 = 0.8$, and $n_{\rm s} = 0.96$. Absolute magnitudes are computed assuming $h=1$.

\section{Data}\label{sec:data}

The data employed in this work were collected by the Kilo Degree Survey (KiDS), a multi-band imaging survey that has mapped $1350$ deg$^2$ of the sky, divided in two equally sized patches, one in the equatorial region and one in the southern hemisphere. The data release we use here (DR4, hereafter KiDS-1000) covers $1006$ deg$^2$ and provides high quality images in the $ugri$ bands, obtained on the VLT Survey Telescope \citep[VST;][]{Capaccioli2012} with the OmegaCAM instrument \citep{Kuijken2011}. By survey design, the best images are provided in the $r$ band, where the mean magnitude limit is $r \sim 25$ ($5\sigma$ in a $2''$ aperture); thus we always refer to the $r$-band images in the rest of this work. Five infrared bands, $ZYJHK_\mathrm{s}$, obtained from the companion VISTA Kilo-degree INfrared Galaxy survey \citep[VIKING;][]{Edge2013} complement the data, allowing for a robust photometric redshift calibration \citep[][]{Wright2019A, Hildebrandt2020, Vakili2023}. 

\subsection{The lens sample} \label{subsec:lens_sample}

The LRG sample is selected from KiDS-1000 using a variation of the \textsc{redMagiC} algorithm \citep{Rozo2016redMaGiC}, as presented in \cite{Vakili2019, Vakili2023}. Details of the sample properties can be found in \citet{Vakili2023}. Here, we summarise the most relevant ones. The sample is selected with a redshift-dependent magnitude cut to ensure a constant comoving number density. The parameter that regulates the selection is $m_{r, \mathrm{pivot}}(z)$, the characteristic $r$-band magnitude of the \cite{Schechter1976} function, assuming a faint-end slope of $\alpha=1$. The resulting luminous-threshold samples are defined by the ratio
\begin{equation}
    \frac{L}{L_{\rm pivot}(z)} = 10^{-0.4 \left( m_r - m_{r, \rm pivot}(z) \right)} \ ,
\end{equation}
where $m_{r,\mathrm{pivot}}(z)$ is evaluated using the \textsc{EzGal} \citep{Mancone2012EzGal} implementation of \citet{BruzualCharlot2003} models, which are galaxy template models generated using a Chabrier IMF, exponentially decreasing star formation and with no dust. \citet{BruzualCharlot2003} used spectral evolution models of stellar populations at a resolution of 3 Å across the whole wavelength range from 3200 to 9500 Å for a wide range of metallicities. Their models incorporate thermally pulsing stars on the asymptotic giant branch. Specifically, $m_{r,\mathrm{pivot}}(z)$ can be evaluated by assuming a Salpeter initial mass function \citep[][]{Chabrier2003}, a solar metallicity ($Z=0.02$), and a single star formation burst at $z=3$.

Two samples are obtained with the aforementioned strategy: a high luminosity ($L/L_{\rm pivot}(z)>1$), sparser sample ($\bar{n}_{\rm g} = 2.5 \times 10^{-4} h^{3} {\rm Mpc}^{-3}$) named the \lum\ sample, and a denser ($\bar{n}_{\rm g} = 10^{-3} h^3 {\rm Mpc}^{-3}$), less luminous one ($L/L_{\rm pivot}(z)>0.5$) as the \dense\ sample. In this work, we followed \citet{Fortuna2021b} and used both samples for our analysis, removing from the \dense\ samples the galaxies that are in common with the \lum\ sample. Because the galaxies are selected imposing constant comoving number densities, the two samples partially overlap in their physical properties, even after the removal of the overlapping galaxies; for a detailed explanation of why this is the case, we refer to \citet[][]{Vakili2023} and  \citet[][]{Fortuna2021b}. We also adopted the same luminosity-binning scheme as \citet[][]{Fortuna2021b}, with some minor variations, described in Sect.~\ref{sec:lensing_results}.  

We quantified the scatter in the photometric-spectroscopic redshift relation using the scaled median absolute deviation of $(z_{\rm phot} - z_{\rm spec})/(1+z_{\rm spec})$. This increases with redshift and it is tighter for the \lum\ sample. In particular, we found $\sigma_z = 0.0139$ for the \lum\ sample and $\sigma_z =0.0146$ for the \dense\ sample.  This effect is also responsible for some overlap of the galaxy properties between the two samples, even when removing the overlapping galaxies.

The stellar masses and absolute magnitudes are obtained via \textsc{LePhare} \citep{Arnouts1999, Ilbert2006, Arnouts2011Lephare}, assuming the  stellar  population  synthesis model from \citet[][]{BruzualCharlot2003}, the \citet{Chabrier2003} initial mass function, and \citet{Calzetti1994} dust-extinction law. We used the \texttt{MASS\_BEST} output from \textsc{LePhare} as our best estimate of the luminous mass to use for the point mass approximation (see Sect.~\ref{sec:model}). When computing the mean stellar mass per bin, we remove the galaxies for which the masses estimated as \texttt{MASS\_MED} are flagged as bad (-99), indicating that the best fit was performed by a non-galaxy template. These galaxies are, however, used in the measurements, as our main focus is the luminosity-to-halo mass relation and the luminosity is robustly measured. We note that the change in the average mass is at the sub-percent level.

We corrected the absolute magnitudes for the passive evolution of the stellar population ($e-$correction). We used \textsc{EzGal} to compute it and follow the setup used in \citet[][]{Vakili2023} to identify the limiting magnitude for the selection of the LRG candidates as described above.

\subsection{The source sample}
\label{subsec:source_sample}

\begin{figure}
    \centering
    \includegraphics[width=\columnwidth]{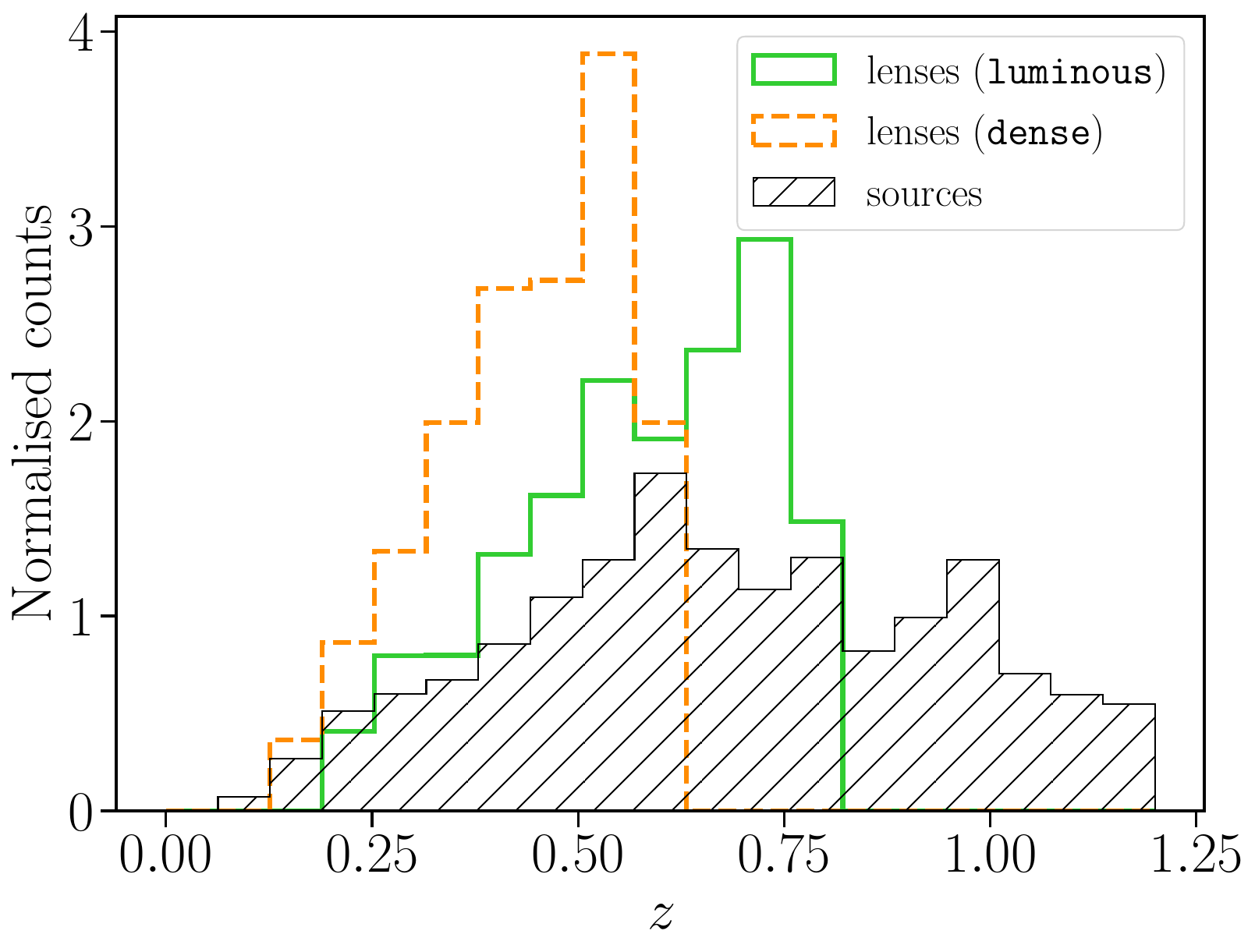}
    \caption{Photometric redshift distributions of the source and lens samples employed in the analysis. }
    \label{fig:z_hist_source_vs_lenses}
\end{figure}

The shapes of our galaxies are computed via a self-calibrating version of \lensfit\ \citep{Miller2007, Miller2013, FenechConti2017, Kannawadi2019, Giblin2021}.\footnote{The catalogue is publicly available at \url{http://kids.strw.leidenuniv.nl/DR4/KiDS-1000_shearcatalogue.php}} This is a model-based algorithm that provides a measure of the ellipticity by fitting a PSF-convolved two-component bulge and disc model of a galaxy. It returns the ellipticity components $\epsilon = \epsilon_1 + i \epsilon_2$, with $|\epsilon| = (a-b)/(a+b)$, where $a,b$ are, respectively, the major and minor axis. For each galaxy, the method returns also a weight, $w_{\rm s}$, which accounts for the increase and decrease in the signal-to-noise ratio (S/N) due to the relative orientation of the galaxy with respect to the PSF and the overall S/N. 
We note that with this definition, the average ellipticity is an estimator of the shear, $\langle \epsilon \rangle = \gamma$. 

The KiDS-1000 shear catalogue benefits from the improvement in the PSF treatment due to the available information provided by the \textit{Gaia} mission \citep[][]{Brown2018GaiaDR2}. The shears are also calibrated to account for the multiplicative bias ($m-$bias) that arises as a further correlation between shear systematics: this can be calibrated employing high fidelity image simulations based on deep images from the Cosmic Evolution Survey \citep[COSMOS,][]{Scoville2007}. We  discuss the $m-$bias again in Sect.~\ref{sec:galaxy_galaxy_lensing}, where we explain how we applied it to our lensing measurements. A full description of the catalogue and the systematic tests can be found in \citet[][]{Giblin2021}, while more details on the strategy to calibrate residual biases are described in \citet[][]{Kannawadi2019}. 

The distributions of our photometric redshifts are illustrated in Fig.~\ref{fig:z_hist_source_vs_lenses}. These redshifts of the source galaxies are estimated based on deep spectroscopic catalogues that cover a sub-sample of the galaxies: these were then re-weighted using a self-organised map \citep[SOM,][]{Wright2020} to resemble the KiDS-1000 sample. Only sources with matches in the calibration sample in the SOM entered our final sample. The method was also validated by using a clustering$-z$ algorithm. Details on the photometric redshift calibration can be found in \citet[][]{Hildebrandt2021photoz}.
We restricted our analysis to source galaxies in the redshift range $0.1<\zs<1.2$, based on the available calibration of their photometric redshift via the SOM, where $\zs$ indicates the photometric redshift of the source galaxy. When computing the GGL signal, we only consider source galaxies at higher redshift than the lens, and with a minimum redshift separation of 0.2, $z_{\rm s} \geq 0.2 + z_{\rm l}$, where the subscripts s and l refer  to the sources and the lens, respectively.

\section{Measuring the signal}\label{sec:galaxy_galaxy_lensing}

The GGL signal is quantified by the tangential distortion in the shapes of background galaxies (sources)
induced by the mass distribution of the foreground galaxies (lenses) along the line of sight. As the distortion for a single lens galaxy is small and we lack information on the intrinsic shape of the background galaxy, we performed a statistical analysis of the signal encoded by a large number of lens-source galaxy pairs and measured the mean tangential shear around each lens as a function of lens-source galaxy projected separation, $\langle \gamma_{\rm t} (\rp) \rangle$. This is a direct measure of the enclosed mass, as we  explain later in this work. We note that the S/N of the lensing signal around individual lenses is too low to be detected, thus, we averaged the signals of an ensemble of lenses. Here, we are implicitly assuming the weak lensing regime, so that the effective shear of a background galaxy can be approximated by the sum of the shears of the individual galaxies in the foreground. We measure the signal of both the \dense\ and \lum\ samples in bins of luminosity.

For each source-lens pair, we measure the tangential component of the ellipticity: indicating with $\phi$ the angle between the $x-$axis and the lens-source separation vector. Using the ellipticity definition introduced in Sect.~\ref{subsec:source_sample}, we have
\begin{equation}
    \begin{bmatrix}\epsilon_{\rm t} \\ \epsilon_{\times} \end{bmatrix} = \begin{bmatrix} -\cos(2\phi) & -\sin(2\phi) \\ \sin(2\phi) & -\cos(2\phi)  \end{bmatrix} \begin{bmatrix}\epsilon_{1} \\ \epsilon_{2} \end{bmatrix} \ .
\end{equation}
Here, $\epsilon_{\times}$ is the cross-component of the ellipticity, which corresponds to a rotation of $45\deg$ from the tangential direction. The cross-component is an important test of residual systematics and in our measurements, we always ensure that the cross-component is compatible with noise. 

The ensemble average of the ellipticities of all the sources, which we recall is an estimator of the shear, at a projected separation, $\rp$, from the lens is directly related to the amount of matter that we observe around a galaxy. This is quantified by the excess surface mass density (ESD) profile:
\begin{equation} \label{eq:esd}
     \Delta \Sigma (\rp) = \bar{\Sigma}(<\rp) - \Sigma (\rp)  = \gamma_{\rm t} (\rp) \ \Sigma_{\rm crit}  \ .
\end{equation}
The ESD is thus defined as the difference between the mean projected surface mass density enclosed in a projected radius, $\rp$, and the surface mass density at $\rp$. The critical surface mass density is a geometrical factor defined as
\begin{equation} \label{eq:sigma_crit}
    \Sigma_{\rm crit} = \frac{c^2}{4 \pi G (1+\zl)^2} \frac{D(\zs)}{D(\zl) D(\zl,\zs)} \ ,
\end{equation}
where the factor $(1+\zl)^{2}$ at the denominator accounts for our use of comoving units \citep[see also Appendix C of][for a discussion of this term]{Dvornik2018}. Here, $\zl$ ($\zs$) is the redshift of the lens (source) galaxy, while $D(\zl)$, $D(\zs),$ and $D(\zl, \zs)$ are the angular diameter distance to the lens, to the source, and between the lens and the source galaxies, respectively.

As we rely on photometric redshifts, we need to integrate Eq.~\eqref{eq:sigma_crit} for the redshift probability distributions of the source sample, $n(\zs')$, and the individual redshift probability distribution of each lens, $p(\zl'|\zl)$. This provides an effective estimate of $\Sigma_{\rm crit}$ as a function of the lens photometric redshifts:
\begin{multline}
    \Sigma_{\rm crit, eff}^{-1} (\zl) = \frac{4 \pi G}{c^2} \int_0^{\infty} (1+\zl')^2 D(\zl') \left( \int_{\zl}^{\infty}  \frac{D(\zl',\zs')}{D(\zs')} n(\zs') \dd \zs' \right) \\
    p(\zl'|\zl) \dd \zl' \ .
\end{multline}

To model $p(\zl'|\zl)$ we used a Gaussian centred on the photometric redshift of the given lens and with standard deviation given by the value of $\sigma_z$ associated with the specific sample, as discussed in Sect.~\ref{subsec:lens_sample}. The redshift probability distribution of the source sample, $n(\zs')$, is instead
obtained from the SOM as described in Sect.~\ref{subsec:source_sample}.

The lensing signal decreases as the distance between the lenses and the sources decreases (due to the $D(\zl, \zs)$ in $\Sigma^{-1}_{\rm crit}$). In our case, while the lenses span a large range in redshift ($0.15 < \zl < 0.8$), the signal is limited by the source sample, for which we have robust redshift estimates only up to $\zs = 1.2$ \citep[][]{Wright2020}. This means that our lensing efficiency peaks around $z_{\rm ls}\sim0.3$ and rapidly decreases as we approach high redshifts.

To each lens-source pair, we also assigned a weight determined by three components: a weight associated with the source sample, $w_{\rm s}$, which down-weights the shears of the galaxies with low S/N, and that corresponds to the \lensfit\ weight reported as \texttt{weight} in the KiDS-1000 shear catalogue (see Sect.~\ref{subsec:source_sample}); a weight associated with the lens galaxies, which is designed to remove residual correlations between the spatial galaxy number density and the survey observing conditions \citep{Vakili2023}; and a geometric term that down-weights lens-source pairs that are close in redshift, given by the square of the inverse critical mass surface density:
\begin{equation}\label{eq:lensing_efficiency}
    w_{\rm ls, eff} = w_{\rm s} w_{\rm l} \left( \Sigma_{\rm crit, eff}^{-1} \right)_{\rm l}^2 \ ,
\end{equation}
where we indicated with a subscript $\rm l$ the lens redshift dependence of the effective critical surface density.

Our estimator (here indicated with a hat) for the excess surface mass density thus reads:
\begin{equation}\label{eq:esd_ls_estimator}
    \hat{\esd}_{\rm ls} (\rp) = \left[ \frac{\sum_{\rm ls} w_{\rm ls, eff} \  \epsilon_{\rm t,s} \ (\Sigma_{\rm crit, eff})_{\rm l}}{\sum_{\rm ls} w_{\rm ls, eff} } \right] \frac{1}{1+\bar{m}} \Bigg|_{\rp} \ ,
\end{equation}
where we have included an average correction to the galaxy shear obtained from dedicated simulations, which quantifies the residual multiplicative bias in the estimate shear due to the presence of noise and blending in the images. The $m-$bias is a function of redshift: here, we rely on the calibration presented in \citet{Kannawadi2019} and evaluate it in narrow redshift slices and weight it by $w' = w_{\rm s} D(\zl, \zs)/D(\zs)$:
\begin{equation}
    \bar{m} = \frac{\sum_i w_i' m_i}{\sum_i w_i'}
,\end{equation}
where $i$ is the $i-$th redshift slice. The $m$ bias measured in our samples goes from $-0.01$ to $-0.03$.

\subsection{Contamination from physically associated galaxies: Boost factor and IA}

Because galaxies tend to cluster and the clustering is a function of galaxy separation, there is an over-density of sources that are physically associated with the lens. This has two implications: on the one hand, these galaxies are not lensed, diluting the GGL signal at small scales; on the other hand, because these galaxies experience the local tidal field, some of them are intrinsically aligned towards the lens; that is, they are opposite to the lensing signal, further suppressing the signal. The former effect can be accounted by comparing the weighted number of pairs between the lens and source sample and the weighted number of pairs that a random distribution of lenses forms with the source sample, as a function of the projected separation, $\rp$ \citep{Sheldon2004, Mandelbaum2005}. This term is typically referred as boost factor:
\begin{equation}\label{eq:boost_factor}
   B(\rp) = \frac{\sum_{\rm r} w_{\rm r}}{\sum_{\rm l} w_{\rm l}} \frac{\sum_{\rm ls} w_{\rm ls, eff}}{\sum_{\rm rs} w_{\rm rs, eff}} \Bigg|_{\rp}
,\end{equation}
where $ w_{\rm rs, eff} = w_{\rm s} \left( \Sigma_{\rm crit, eff}^{-1} \right)^2$ and a weight associated with randoms $w_{\rm r}$ is set to 1. 

Both the lensing dilution from unlensed galaxies and the negative contribution from IA can be removed by selecting only source galaxies that have separations larger than $z_{\epsilon_l}$ from the lens, with $\zs - \zl = z_{\epsilon_l}$ \citep[][]{Leauthaud2017LensingIsLow}. Although we also applied the boost factor, we made use of this cut when measuring the lensing signal to ensure that any contamination is low and we adopted $ z_{\epsilon_l}=0.2$. 

\subsection{Random subtraction}

On top of the correction discussed in the previous section, we followed \citet{Singh2017} and subtracted the signal around random points from the 
lensing signal. This ensures that residual additive biases, introduced by the survey edges and by the presence of masks, would be removed from our measurement. The random signal is obtained in exact analogy to Eq.~\eqref{eq:esd_ls_estimator}, but measuring the lensing signal around points uniformly distributed over the survey footprint with removal of the masked regions. The final estimator is thus given by
\begin{equation} \label{eq:final_esd_estimator}
    \hat{\esd}(\rp) = B(\rp) \hat{\esd}_{\rm ls}(\rp) - \hat{\esd}_{\rm rs}(\rp) \ .
\end{equation}

\subsection{Estimator for the IA signal}
\label{subsec:extracting_IA_signal}

We are, however, also interested in measuring the IA signal of the source galaxies around the lenses. To this end, we only selected galaxies within a small redshift separation from the lenses. We chose $\Delta z \equiv |\zl-\zs|<z_{\epsilon}$ with $z_{\epsilon}=0.06$. The computation is analogous to the case of lensing, (Eq.~\ref{eq:final_esd_estimator}): we labelled the resulting signal as $\esd_{\Delta z}$ to denote the redshift range used for this measurement. The signal measured in this way is still affected by the lensing contamination, which can be removed using the lensing signal measured for the `lensing' sample. 
When focussing on physically associated galaxies,  it is crucial to correctly account for the boost factor in such cases. 

The excess of lens-source pairs after the random subtraction described in the previous section is then the obtained IA signal (as this is what remains due to the clustered galaxies): the average critical surface density of the excess pairs,  $\langle \Sigma_{\rm crit, eff} \rangle_{\rm ex}$, thus becomes:
\begin{equation}
    \langle \Sigma_{\rm crit, eff} \rangle_{\rm ex} = \frac{\sum_{\rm ls} w_{\rm ls, eff} \Sigma_{\rm crit, eff}^{\rm (ls)} - \sum_{\rm rs} w_{\rm rs, eff} \Sigma_{\rm crit, eff}^{\rm (rs)}}{\sum_{\rm ls} w_{\rm ls, eff} - \sum_{\rm rs} w_{\rm rs, eff}} \ .
\end{equation}
Finally, the IA estimator is \citep[][]{Blazek2012}:
\begin{equation} \label{eq:ia_estimator}
    \hat{\gamma}_{\rm IA} (\rp)= \frac{\hat{\esd}_{\Delta z} - \hat{\esd}_{\rm lens}}{(B_{\Delta z} -1) \langle \Sigma_{\rm crit, eff}^{(\Delta z)} \rangle_{\rm ex} - (B_{\rm lens} -1) \langle \Sigma_{\rm crit, eff}^{(\rm lens)} \rangle_{\rm ex}} \Bigg|_{\rp} \ .
\end{equation}

\citet[][]{Leonard2018} presented an improved version of this estimator, which exploits the scale dependence of IA to better separate it from lensing. This requires multiple measurements of the source galaxy shapes, obtained with different shape estimates that weigh different galaxy scales differently. Here, we do not investigate this possibility, mainly motivated by the results in \citet[][]{Georgiou2019b}, who used very high S/N measurements and found that the scale dependence of IA is mainly limited to red satellites. The gain is therefore expected to be minimal in a mixed sample.

\section{Modelling the signal with the halo model}\label{sec:model}

The GGL signal captures the projected two-point correlation function between a galaxy position and a galaxy shear, which, in turn, is a measure of the three-dimensional correlation between matter and density distributions. Because lensing is sensitive to density contrasts, in practise, we measured the difference between the projected mass density at a certain radius and the average mass density contained in that radius (Eq.~\eqref{eq:esd}). 

In order to model the signal, we need to provide an analytical expression for the projected surface mass density around galaxies. This is related to the three-dimensional (3D) correlation function via a projection integral. In the distant observer approximation, it can be expressed as an Abel transform:
\begin{equation} \label{eq:sigma_rp}
    \Sigma (\rp) = 2 \bar{\rho}_m \int_{\rp}^{\infty} \xi_{\delta \rm g} (r) \frac{r \ \dd r}{\sqrt{r^2 - \rp^2}} \ .
\end{equation}
Here $\xi_{\delta \rm g}(r)$ is the correlation between the galaxy and the matter density contrast, $\langle \delta(\mathbf{x}) \delta_{\rm g}(\mathbf{x+r}) \rangle$, with $\delta(\mathbf{x}) = (\rho_{\rm X}(\mathbf{x})-\bar{\rho}_{\rm X})/\bar{\rho}_{\rm X}$, $X \in \{\delta, \rm g\}$.  In the following, we will always use the short-notation $\delta$ to indicate the dark matter and ${\rm g}$ for the galaxy. Since galaxies form inside dark matter halos, the halo model is a natural framework to describe the matter-galaxy correlation function, $\xi_{\delta \rm g}(r)$. We present this formalism in the next section. 

The projected mass contained within the radius $\rp$ can be written as
\begin{equation} \label{eq:sigma_lrp}
    \bar{\Sigma}(<\rp) = \frac{2}{\rp^2} \int_{0}^{\rp} \Sigma (R') R' dR'
,\end{equation}
and from Eq.~\eqref{eq:esd} we can recover the ESD.

While this formalism strictly describes only the lensing effect due to the dark matter distribution, we also included the GGL due to the baryonic mass of the galaxy, which is modelled as a point mass (pm) here and added to the predicted ESD signal obtained from the halo model [Eq.~\eqref{eq:esd}],
\begin{equation}
    \Delta \Sigma_{\rm pm} = \frac{\langle M_* \rangle}{\pi r_p^2} \ .
\end{equation}
Here, $\langle M_* \rangle$ is the median stellar mass for each \texttt{dense} and \texttt{luminous} samples.
This approximation is motivated by the fact that the minimum scale we are probing is $r_p= 60 \ h^{-1} {\rm kpc}$, much larger than the physical extent of the stellar component of a galaxy. 

We evaluated the model at the effective redshift of the lenses, $z_{\rm eff}$, given by the weighted mean redshift of the lenses, where the weight is given by the lensing efficiency in Eq.~\eqref{eq:lensing_efficiency}. For the \lum\ sample, this corresponds to $z_{\rm eff} = 0.406$, while for the \dense\ sample  $z_{\rm eff} = 0.368$.

\subsection{The halo model}\label{subsec:halo_model}

The halo model \citep[][for a review]{Seljak2000, MaFry2000, vandenBosch2013-PaperI, CooraySheth2002, 2023Asgari_HaloModelReview} is a well-established formalism for predicting and interpreting the clustering and lensing statistics of galaxies and dark matter. The key idea behind the halo model is that the mass of the halo is the fundamental property that drives halo clustering statistics. It assumes that all dark matter in the Universe is bound in halos and that dark matter halos are fully described by a universal density profile. 

The formalism is based on a set of ingredients: a density profile for the dark matter distribution; a halo mass function that provides a prescription of how many halos populate a given comoving volume at a given redshift; and a halo bias function, which quantifies the bias of the bounded halos with respect to the underlying matter distribution.  Subsequently, galaxies can be included into the formalism through a prescription that provides the way galaxies occupy dark matter halos. The halo occupation distribution (HOD), is a convenient way of doing that, assigning the number of galaxies $N_{\rm g}$ per given halo of mass $M$, $\langle N_{\rm g}|M \rangle$ \citep{Kauffmann1999, Benson2000, Berlind2002, Zheng2005}. We discuss  the model we adopt for the HOD in greater detail in Sect.~\ref{subsec:hod}.

We define dark matter halos as spheres with an average density of 200 times the background density today, $\rho_{\rm h} = 200 \rho_{\rm m}$. We assume that the dark matter in a halo is spatially distributed following the Navarro-Frenk-White profile \citep[NFW, ][]{NFW1996}, with a concentration-mass relation from \citet{Duffy2008}. We also assume that satellite galaxies are spatially unbiased with respect to the dark matter particles; namely, that their spatial distribution is described by $\rho_{\rm s}(r,M) = \rho_{\rm h}(r,M) \equiv M u(r,M)$, with $u(r,M)$ the normalised density profile of dark matter and $r$ the distance from the centre of the halo. We allow central galaxies to have a different amplitude of the concentration-mass relation, which we parametrise as a free pre-factor, $f_{\mathrm{conc}}$. For the halo mass function, $n(M)$, as for the halo bias function, $b_h(M)$, we adopted the functions presented in \citet{Tinker2010}. We explicitly force the halo bias to be normalised at each redshift: the normalisation is obtained by integrating the halo bias function over a large range of masses ($10^{2}-10^{18} \ h^{-1} M_{\odot}$). 
We do not model the off-centering of central galaxies; that is, we always assume that they sit at the centre of their halo.

\subsection{The galaxy-matter power spectrum}

Given a prescription for the HOD (Sect.~\ref{subsec:hod}) and the set of ingredients introduced in Sect.~\ref{subsec:halo_model}, it is possible to build the correlation functions between the matter density field and a continuous galaxy field, as well as their auto-correlations. As such relations involve convolutions, for computational reasons, it is convenient to work in Fourier space and then transform the quantities back to real space. We thus present them in Fourier space, as they are implemented this way in the code. 

One of the main advantages of the halo model is its separate treatment of the correlation that arise between the galaxies and matter within the halo, which leads to the so called one-halo term, and the correlation between those that belong to different halos, namely, the two-halo term. As a general result, the full power spectrum is
\begin{equation}
    P(k,z) = P^{1 \rm h}(k,z) + P^{2 \rm h}(k,z),
\end{equation}
regardless whether we are describing the clustering of galaxies, of dark matter or the matter-galaxy correlation.

In turn, we can split the contributions from central and satellite galaxies and model them individually. We use  `c' to denote the central-galaxy components, with `s' as the terms  sourced by the satellite population and with $\delta$ as those corresponding to matter. We find that any correlation is given by the sum of all of the possible correlations between these terms. In Fourier space, for the case of the galaxy-matter cross power spectrum, this is expressed as:
\begin{equation}
    P_{\rm g \delta}(k,z) = f_{\rm c} P^{1 \rm h}_{\mathrm{c} \delta}(k,z) +  f_{\rm s} P^{1 \rm h}_{\mathrm{s} \delta}(k,z) + f_{\rm c} P^{2 \rm h}_{\mathrm{c} \delta}(k,z) + f_{\rm s} P^{2 \rm h}_{\mathrm{s} \delta}(k,z) \ .
\end{equation}
Here, $f_X$ with $X \in \{ {\rm c,s} \}$ is the fraction of galaxies of a given type entering the correlation. These can be obtained from the galaxy number densities as predicted by the HOD as
\begin{equation}\label{eq:f_g}
    f_X = \frac{n_X}{n_{\rm g}} \ ,
\end{equation}
where
\begin{equation}
    n_X = \int_0^{\infty} \langle N_X | M \rangle \ n(M) \ \dd M \ .
\end{equation}

It is convenient to introduce the field profile functions, $\mathcal{H}_X$, where $X=\{\delta, {\rm c, s}\}$ and are thus associated with a given component:
\begin{equation}
    \mathcal{H}_{\delta} (k, M) = \frac{M}{\rho_{\rm m}} u(k, M) \ ,
\end{equation}
\begin{equation}
    \mathcal{H}_{\rm c} (k, M) = \frac{\langle N_c|M \rangle}{n_g} 
\end{equation}
and
\begin{equation}
    \mathcal{H}_{\rm s} (k, M) = \frac{\langle N_s|M \rangle}{n_g} u(k,M) \ .
\end{equation}

The one-halo and two-halo terms of the power spectrum are then:
\begin{equation}
    P^{1h}_{xy}(k) = \int_0^{\infty} \mathcal{H}_x(k,M) \mathcal{H}_y(k,M) \ n(M) \ \dd M 
\end{equation}
and
\begin{equation}
\begin{split}
    P^{2h}_{xy}(k) = P_{\rm lin}(k)  & \int_0^{\infty} \dd M_1 \ \mathcal{H}_x(k,M_1) b_{\rm h}(M_1) n(M_1) \\ & \times \int_0^{\infty} \dd M_2 \ \mathcal{H}_y(k,M_2) b_{\rm h}(M_2) n(M_2) \ .
\end{split}
\end{equation}

A relevant quantity that we can predict via this formalism is the average mass of bounded halos that host a central galaxy. This is defined as:
\begin{equation}\label{eq:M200}
    \langle M_{\rm eff} \rangle = \frac{1}{n_{\rm c}(z_{\rm eff})} \int \langle N_{\rm c} | M \rangle n(M) M \dd M \ .
\end{equation}

\subsection{The halo occupation distribution} \label{subsec:hod}

Following \citet{Cacciato2009}, we derived the HOD from the conditional luminosity function (CLF) obtained for the SDSS by \citet{Yang2003}. The CLF, d$\Phi(L|M) \dd M$, specifies the average number of galaxies with luminosity in the range $L\pm \dd L/2$ that reside in a halo of mass $M$. Thus, integrating over a certain luminosity bin provides the number of galaxies with a certain luminosity $L\in[L_1,L_2]$ that reside in a halo of mass, $M$,
\begin{equation}
    \langle N_g|M, L_1, L_2 \rangle = \int_{L_1}^{L_2} \Phi(L|M) \text{d} L.
\end{equation}

As in \citet{Cacciato2009}, we split the CLF in two components,
\begin{equation}
    \Phi(L|M) = \Phi_{\text{c}}(L|M) + \Phi_{\text{s}}(L|M),
\end{equation}
where $\Phi_{\text{c}}(L|M)$ is the CLF associated with central galaxies, while $\Phi_{\text{s}}(L|M)$ is the CLF associated with satellite galaxies.

The central galaxy CLF is described by a log-normal function,
\begin{equation}\label{eq:clf_centrals}
    \Phi_\text{c}(L|M) dL = \frac{\log e}{\sqrt{2 \pi}} \exp \left[ - \frac{(\log L - \log L_\text{c})^2}{2 \sigma_\text{c}^2} \right] \frac{\text{d}L}{L},
\end{equation}
where
\begin{equation}\label{eq:Lc(M)}
    L_\text{c} (M) = L_0 \frac{(M/M_1)^{\gamma_1}}{[1+(M/M_1)]^{\gamma_1-\gamma_2}}
\end{equation}
is the mean luminosity of central galaxies in a halo of mass, $M$. Here, $M_1$ is the characteristic mass scale at which $L_\text{c}(M)$ changes its slope, while $L_0$ is the normalisation, with $\gamma_1$ and $\gamma_2$ being the slope at a low-mass end and high-mass end ($L_\text{c} \propto M^{\gamma_1}$ for $M \ll M_1$ and $L_\text{c} \propto M^{\gamma_2}$ for $M \gg M_1$). Then, $\sigma_\text{c}$ is the scatter between luminosity and halo mass. Equation~\eqref{eq:Lc(M)} is one of the key relations we aim to constrain for the LRG sample (see Sect.~\ref{sec:lensing_results}).

Satellite galaxies obey
\begin{equation}\label{eq:clf_satellites}
    \Phi_\text{s} (L|M) = \Phi_{\text{s}}^* \left( \frac{L}{L_{\text{s}}^{*}} \right)^{\alpha_\text{s}+1} \exp \left[ - \left( \frac{L}{L_{\text{s}}^{*}} \right)^2 \right] \frac{\text{d}L }{L},
\end{equation}
where $L_{\text{s}^*}(M)=0.562 L_{\text{c}}(M)$ and 
\begin{equation}
    \log [\Phi_{\text{s}}^{*}(M)] = b_0 + b_1 (\log M_{12}). 
\end{equation}

Here, we have defined $M_{12}$ as $M_{12} = M / 10^{12}\, h^{-1} M_{\odot}$, with the $\alpha_{\mathrm{s}}$, $b_0$, and $b_1$ parameters governing the slope and the normalisation of the Schechter function, respectively. In total, our CLF is described by nine parameters: $\log M_1, \log L_0, \gamma_1, \gamma_2, \sigma_\text{c}, f_{\rm conc}, \alpha_{\rm s}, b_0, b_1$. 

\section{Fitting procedure}\label{sec:fitting_procedure}

We sampled the parameter space via a Markov chain Monte Carlo procedure, using the \textsc{Emcee} \footnote{\url{https://emcee.readthedocs.io/en/stable/}} \citep[][]{Foreman-Mackey2013emcee} sampler. We assumed a Gaussian likelihood of the form:
\begin{equation}
    \mathcal{L} \propto \exp \left[ -\frac{1}{2} \left(\mathbf{M(\theta) - \mathbf{D}}\right)^{\rm T} \mathbf{C}^{-1} \left(\mathbf{M(\theta) - \mathbf{D}}\right) \right] \ ,
\end{equation}
where $\mathbf{D}$ is the data vector, $\mathbf{M(\theta)}$ is the model evaluated for the set of parameters $\mathbf{\theta}$, and $\mathbf{C}^{-1}$ is the inverse of the data covariance matrix. We employed 120 walkers. The priors for the CLF parameters reported in Table~\ref{tab:HODpriors}  are based on previous results in the literature. In particular, while we broadly follow the choice of the priors in \citet{Bilicki2021}, we adopted more informative priors in the following cases: $\sigma_{\rm c}$ has been shown to be tightly constrained by current measurements as investigated in \citet[][see e.g. their Fig. 6]{Cacciato2014}; whereas $\gamma_1$ is expected to be poorly constrained by a luminous sample, such as the LRGs. Here, we follow \citet{Cacciato2014}, but rather than fixing it, we provide an informative prior centred on the best-fit value in \citet[][]{Cacciato2013}; the prior for $\gamma_2$ is typically extremely broad, however, the best-fit values obtained for different samples are all in good agreement and span the range $0.2-2.0$ \citep[][]{Cacciato2013, Cacciato2014, vanUitert2016GAMAgroups, Dvornik2018, Bilicki2021}. Thus, we  restricted the sample to the range of $\mathcal{U}(0,2)$. We also reduce the range of $f_{\rm conc}$ based on some preliminary runs and exclude zero to avoid any non-physical behaviour in the model.

\begin{table}
    \centering
    \caption{Priors adopted in the fit and the corresponding fiducial values.}
    \begin{tabular}{ccc}
    \hline
    \hline
        Parameter & Prior & Fiducial \\
        \hline
$f_{\mathrm{conc}}$ & $[0.1,1]$ & $0.98_{-0.054}^{+0.017}$ (0.999)\\
$\log (L_0/[h^{-2} L_{\odot}])$ & $[7,13]$ & $9.95_{-0.577}^{+0.331}$ (10.357) \\
$\log (M_1/[h^{-1} M_{\odot}])$ & $[9.0, 14.0]$ & $11.66_{-0.434}^{+0.341}$ (12.070) \\
$\gamma_1$ & $\mathcal{N}(3.18, 2)$ & $4.22_{-1.863}^{+1.888}$ (3.146) \\
$\gamma_2$ & $[0,2]$ & $0.42_{-0.140}^{+0.174}$ (0.266)\\
$\sigma_{\mathrm{c}}$ & $[0.1,0.3]$ & $0.13_{-0.022}^{+0.060}$ (0.115)\\
$\alpha_{\mathrm{s}}$ & $\mathcal{N}(-1.1,0.9)$ & $-1.73_{-0.404}^{+0.524}$ (-1.478)\\
$b_0$ & $\mathcal{N}(0,1.5)$ & $-1.36_{-0.456}^{+0.491}$ (-1.607)\\
$b_1$ & $\mathcal{N}(1.5,2.0)$ & $0.86_{-0.359}^{+0.287}$ (1.056)\\
    \hline
    \end{tabular}
    \small
    \tablefoot{Fiducial values are here reported as the median of the marginal posteriors, while the best fit values are reported in brackets. The error bars correspond to the 16th and 84th percentiles. $\mathcal{N}(\mu, \sigma)$ indicates a normal distribution with mean $\mu$ and standard deviation $\sigma$. Parameter $f_{\mathrm{conc}}$ is the normalisation of the concentration-halo mass relation, $L_0$ the normalisation of $L_{\mathrm{c}}(M)$ relation, $M_1$ the characteristic scale where $L_{\mathrm{c}}(M)$ changes the slope from $\gamma_1$ to $\gamma_2$. Furthermore $\sigma_\text{c}$ is the scatter between luminosity and halo mass, while $\alpha_{\mathrm{s}}$, $b_1$, and $b_2$ govern the slope and the normalisation of the Schechter function, respectively.}
    \label{tab:HODpriors}
\end{table}

\section{Constraints on the lens sample properties} \label{sec:lensing_results}

\begin{figure*}
    \centering
    \includegraphics[width=\textwidth]{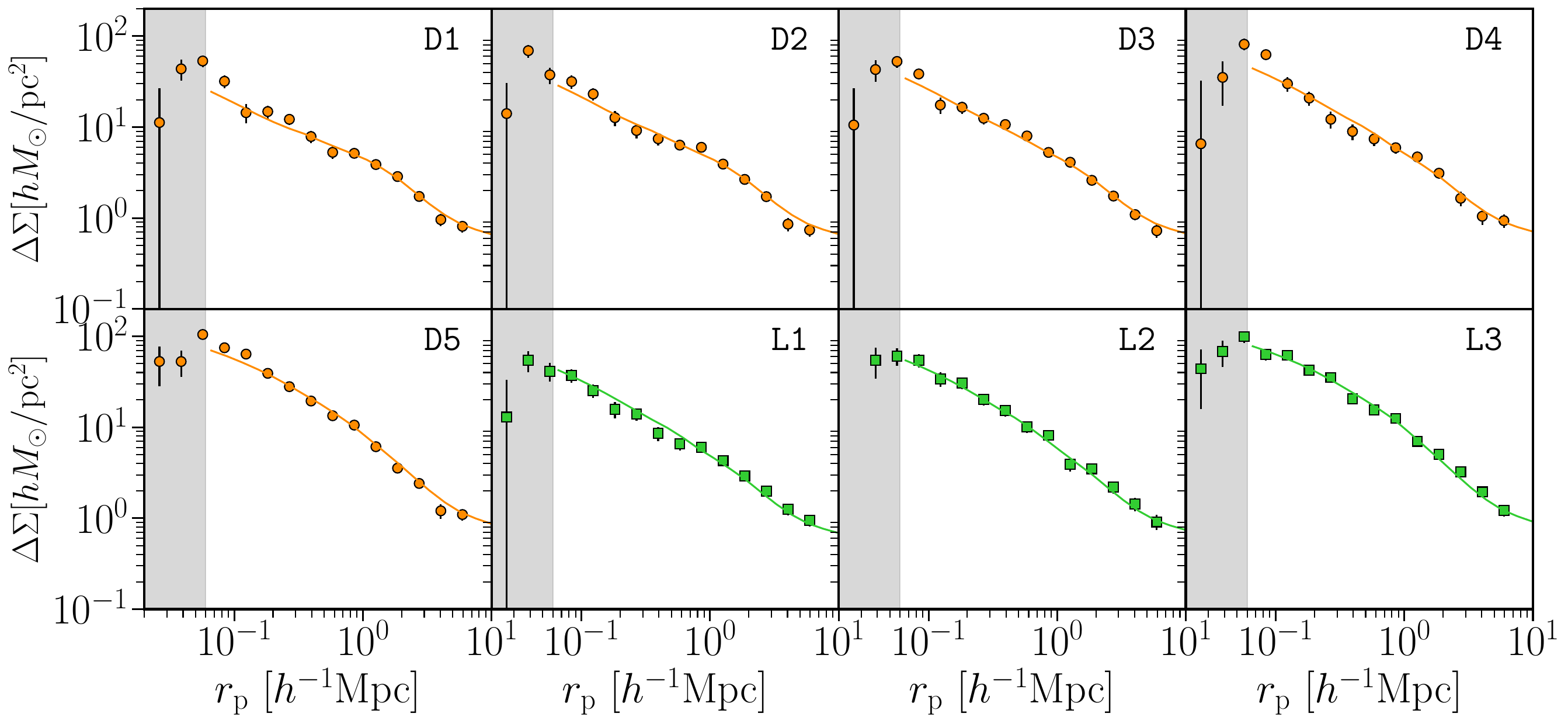}
    \caption{ ESD measurements for the samples listed in Table~\ref{tab:lenses_properties}. We plot the best fit curves (Table~\ref{tab:HODpriors}) on top of the data points. The grey shadowed area is excluded from the fit (see Sect. \ref{sec:lensing_results}). The reduced $\chi^2$ of the fit is $\chi^2_{\rm red} = 1.05$. }
    \label{fig:esd}
\end{figure*}

We measured the ESD signal of the \dense\ and \lum\ samples in bins in luminosity, applying the cuts presented in \citet{Fortuna2021b} and labelled them as \texttt{D1, D2, D3, D4, D5, L1, L2, L3} (Fig. ~\ref{fig:esd}). The \texttt{D1} and \texttt{L1} samples of this study slightly differ from those in \citet{Fortuna2021b} because of the removal of the galaxies that reside in the tails of the distributions. The tail is due to the photo-z scatter, such that a $m_r(z)$ cut does not translate into a sharp cut in absolute magnitudes. We also removed part of the high-$L$, with a cut at $M_r - 5 \log(h) = -22.6$. Removing the tail is crucial for the correct interpretation of the luminosity distributions by the model: the CLF modelling assumed here is designed for volume-complete samples. Given the lack of a selection function in the model, the long faint-tail would be populated by a large number of faint galaxies, as predicted by the modified Schechter function in Eq.~\eqref{eq:clf_satellites}. We explore how well our model can capture the luminosity distribution of our samples in Appendix~\ref{A:L_distr}.

The properties of the lens samples are reported in Table~\ref{tab:lenses_properties}. We jointly fit all the samples with a single model and found a unique set of CLF parameters, which we report in Table~\ref{tab:HODpriors}. The reduced $\chi^2$ is $1.05$. When performing the fit, we do not consider the measurements at $r_p<0.6 h^{-1} \rm Mpc$: at that separation, galaxies start being blended from the light of the LRGs, contaminating the shape measurement \citep[see e.g.][]{Sifon2015}{}{}.

The best-fit CLF parameters agree within the error bars with the best-fit parameters of the red population of the KiDS-Bright sample \citep[][]{Bilicki2021}.
This is a sign that the two samples, despite having been selected with different cuts, are characterised by similar scaling relations. It is, however, surprising that the stellar-to-halo mass relation of the red galaxies of the Bright sample has a very similar scaling to the luminosity-to-halo mass relation of the LRG sample (see Sect.~\ref{subsec:L-M_relation}). We interpret this result as a consequence of the observed luminosity-to-stellar mass-relation for the LRGs, which is close to unity (see Tab.~\ref{tab:lenses_properties}). 

\subsection{Luminosity-halo mass relation}\label{subsec:L-M_relation}

\begin{figure*}
    \centering
    \includegraphics[width=\textwidth]{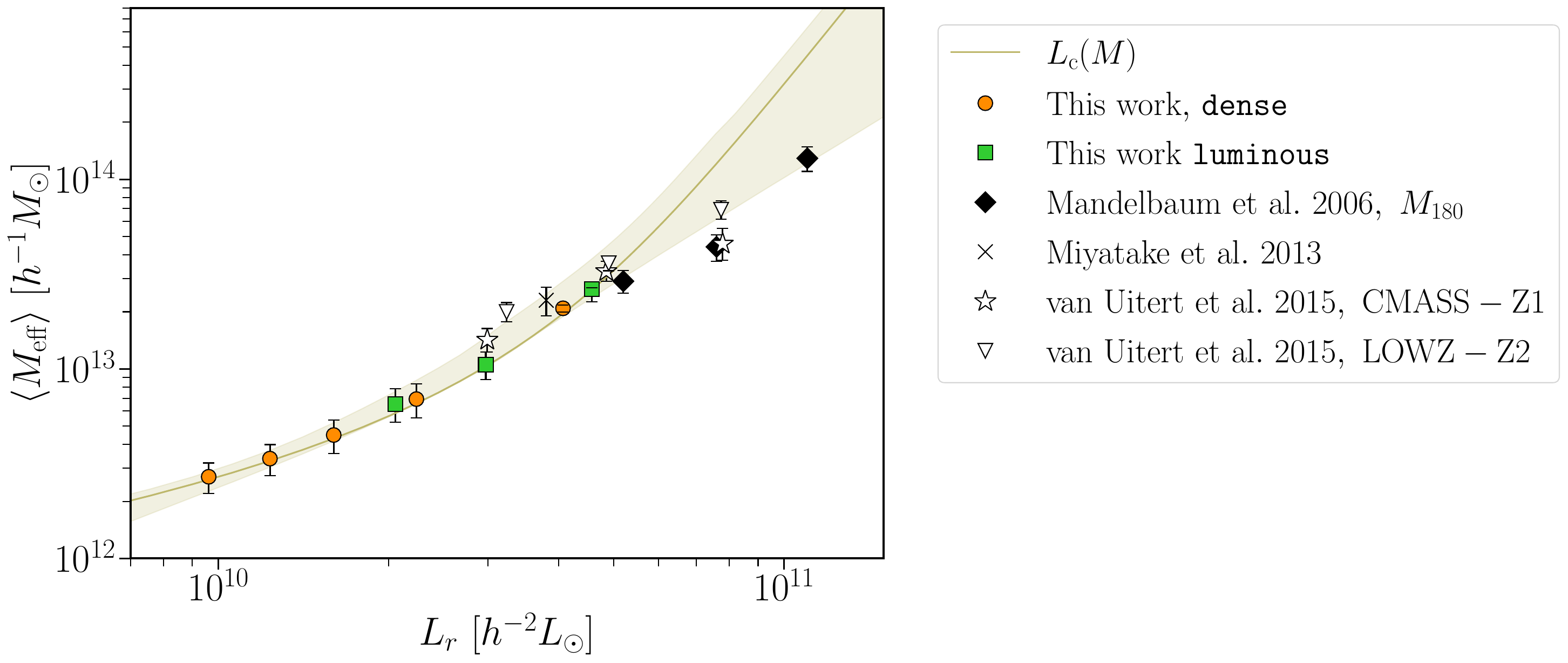}
    \caption{Luminosity-halo mass relation for the LRG sample (\dense\ orange circles; \lum\: green squares), compared with similar measurements from different samples in the literature. The solid line shows the $L_{\rm c}(M)$ relation predicted by our model. We caution that the points from the literature use somewhat different mass definitions and, thus, the comparison is intended for qualitative purposes only. }
    \label{fig:Meff}
\end{figure*}

\begin{table*}
    \centering
     \caption{Properties of the lens samples.}
    \begin{tabular}{cccccc}
    \hline
    \hline
    Sample & $\langle z \rangle $ & $\langle \log L [h^{-2} L_{\odot}] \rangle$ & $\langle \log M_* [h^{-2} M_{\odot}] \rangle$ & $\langle \log M_{\rm eff} [h^{-1} M_{\odot}] \rangle$ & $f_{\rm s}$ \\
    \hline
    \texttt{D1}     &  0.41 & 10.01  & 10.25  &  $ 12.43 \pm 0.05 $  & $0.29 \pm 0.05 $\\
    \texttt{D2}     &  0.42 & 10.10 
    & 10.35  &  $ 12.53 \pm 0.06 $  &   $0.27 \pm 0.04 $\\
    \texttt{D3}     &  0.43 &  10.21  & 10.46  &  $ 12.65 \pm 0.07 $  & $0.24 \pm 0.03 $\\
    \texttt{D4}     &  0.45 &  10.35  & 10.62  &  $ 12.84 \pm 0.08 $  & $0.20 \pm 0.02 $\\
    \texttt{D5}     &  0.45 &  10.59  & 10.84  &  $ 13.32 \pm 0.03 $   & $0.13 \pm 0.01 $ \\
    \texttt{L1}     &  0.53 &  10.33  &  10.63 &  $ 12.81 \pm 0.08 $ & $0.19 \pm 0.02$\\
    \texttt{L2}     &  0.55 &  10.48  &  10.77 &  $ 13.02 \pm 0.07 $ & $0.16 \pm 0.02 $\\
    \texttt{L3}     &  0.56 & 10.65  & 10.94  &  $ 13.42^{+0.02}_{-0.05}$  & $0.11 \pm 0.02$\\
    \hline
    \end{tabular}
    \small
    \tablefoot{$\langle M_{\rm eff} \rangle$ and $f_{\rm s}$ are derived from the set of CLF parameters that maximise the likelihood. The error bars correspond to the 16th and 84th percentiles.}
    \label{tab:lenses_properties}
\end{table*}

For each sample, we derived the corresponding average halo mass, $\langle M_{\rm eff} \rangle$ (Eq.~\ref{eq:M200}). These values are reported in Table~\ref{tab:lenses_properties}. At intermediate luminosity (\texttt{D5, L2 L3}), our results are in good agreement with previous studies \citep[][]{Mandelbaum2006a, Miyatake2015, vanUitert2015}, although at high luminosity the extrapolation of our best-fit curve is above the measurements from literature. This is illustrated in Fig.~\ref{fig:Meff}, where we also plot the luminosity-halo mass relation for the central galaxies as predicted in Eq.~\eqref{eq:Lc(M)}. Here, we only show the point from \citet{vanUitert2015} which are the closest to the effective redshift of our samples. We also note that \citet{Mandelbaum2006a} used a different definition of halo mass; thus, the comparison has to be considered only qualitative. However, our samples are overall fainter than ones from \citet{Mandelbaum2006a}, providing an extension to the $L-M$ towards lower luminosities. Given that the fraction of satellites is overall low (see Table~\ref{tab:lenses_properties}), a qualitative agreement between the data and the curve is expected and the curve provides a useful comparison for simulations and the galaxy properties in mock catalogues.

\subsection{IA dependence on halo mass}

\begin{figure}
    \centering
    \includegraphics[width=\columnwidth]{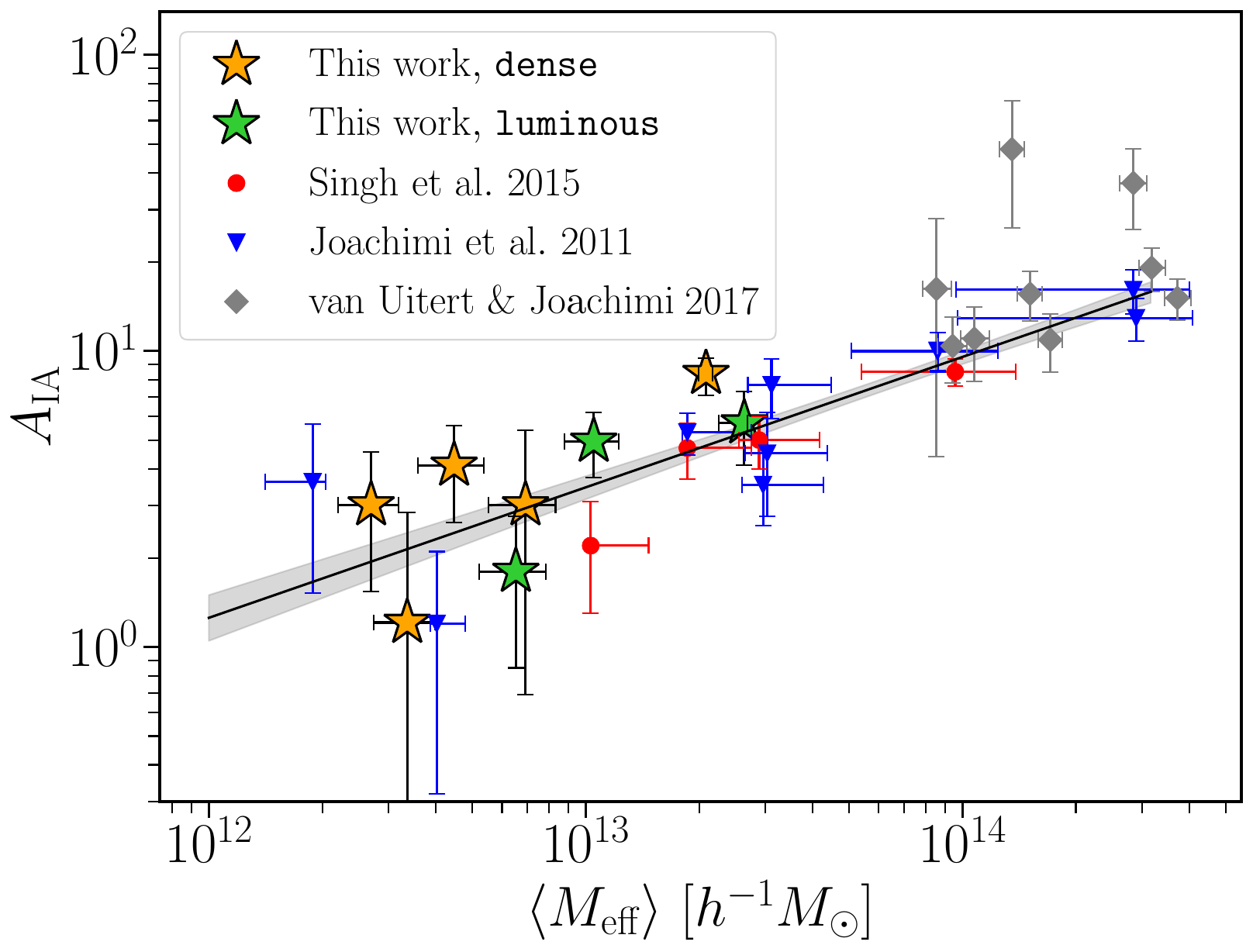}
    \caption{Dependence of IA on halo mass for different IA measurements. The halo masses, $\langle M_{\rm eff} \rangle$, were obtained as described in the text. We indicate our measurements with star markers (orange: \dense\ sample, green: \lum); the other data points are taken from the literature. The solid black line shows the best fit curve of the $A_{\rm IA}(M)$ relation described by Eq.~\eqref{eq:A_mass}, while the shaded area delimits the $68\%$ confidence region.}
    \label{fig:IA_Meff}
\end{figure}

To explore the implications of the observed luminosity-to-halo mass relation for the IA signal of the LRGs, we used the measurements in \citet{Fortuna2021b} and placed them into context, using the estimates of the halo masses obtained in the previous section.
\citet{Fortuna2021b} used a non-linear linear alignment model \citep[][]{Hirata2004, BridleKing2007}, adapted to account for the photometric redshift uncertainty \citep[][]{Joachimi2011b}, to fit the IA signal at large scales ($\rp > 6 \mpch$). The best-fit IA amplitudes, $A_\mathrm{IA}$, of the different sub-samples are shown in Fig.~\ref{fig:IA_Meff} as orange (\dense) and green (\lum) stars.

When combined with previous measurements reported in the literature, \citet{Fortuna2021b} found that the IA dependence on luminosity of the red galaxies can be aptly described by a double power law, with a break at $L_r\lesssim 3.2 \times 10^{10} h^{-2} L_{r,\odot}$, although the data show a large scatter at low-$L$. We are now in a position to investigate whether this dependence was primarily sourced by the halo mass or if the relation is more complex and requires the addition of secondary galaxy properties. Interestingly, our measurements of $L(M)$ lie among the transition between the two regimes of the power law in Eq.~\eqref{eq:Lc(M)}. This means that the double power law in the IA-$L$ plane reflects at least partially the double power law in the $L-M$ plane. We explore this further by showing the IA amplitudes from \citet{Fortuna2021b} as a function of weak lensing mass in Fig.~\ref{fig:IA_Meff}. Although the scatter is too large to draw definitive conclusions, we note that the overall trend matches a single power law. To obtain a more complete picture, we also added the measurements from \citet[][]{Joachimi2011b} (MegaZ, SDSS LRGs, the L3 and L4 samples from the SDSS) and \citet[][]{Singh2015} (LOWZ) to Fig.~\ref{fig:IA_Meff}, which are based on LRGs and thus can safely be assumed to be mainly centrals. For this reason, we decided to omit the measurements from \citet[][]{Johnston2019}, which are known to have a larger fraction of satellites. To do so, we converted the luminosity of each sample into an estimate of their halo mass, via the relation displayed in Fig.~\ref{fig:Meff}. \citet[][]{Singh2015} provide estimates of the halo masses of their samples, but those are based on a different definition of halo mass and so, for ease of comparison, we decided to use our scaling relation. Since \citet[][]{vanUitert2017} did not provide the luminosity of the clusters, but used the same definition of halo mass, we decided to use their halo mass estimate. We fit all the measurements in Fig.~\ref{fig:IA_Meff} with a single power law of the form:
\begin{equation}\label{eq:A_mass}
    A_{\rm IA}(M) = A \left( \frac{M}{M_0}  \right)^{\beta_M} \ ,
\end{equation}
with $M_0 = 10^{13.5} h^{-1} M_{\odot}$. We find a best-fit amplitude of $A = 5.74^{+0.32}_{-0.32}$ and slope $\beta_M = 0.44^{+0.04}_{-0.04}$.
The reduced $\chi^2$ is 1.64 for 21 degrees of freedom. We stress, however, that some of the masses associated with  these measurements lie beyond the range constrained by our data,  thus, they are an extrapolation. This is relevant because a small variation of the slope becomes significant at high$-L$. We test the impact of this by replacing our $L-M$ relation with the one in \citet[][]{vanUitert2015} for the high-mass points (LOWZ, MegaZ and SDSS LRGs).\footnote{The relation is a single power law in the form $M_{\rm eff} = M_{0,L}\,(L/L_{0})^{\beta_{L}}$, using a pivot luminosity of $L_{0} = 10^{11} h^{-2}_{70} L_{\odot}$.} This relation was also adopted in \citet[][]{Piras2018} and includes a redshift dependence, which we do not consider in our model. When repeating the fit with this new set of data, we still find comparable results within the model uncertainty; although we do notice an improvement in the reduced $\chi^2$, which in this case is: 1.19. We also tested the impact of the measurements from galaxy clusters \citep{vanUitert2017} on our fit: the interpretation of their IA signal is complicated by the fact that their shapes were obtained with a different technique than for the LRG samples \citep[][]{vanUitert2017}. We thus removed those measurements from our data vector and repeated the fit: the result of this test is compatible with our fiducial setup.

\section{Constraints on the IA of the source sample}\label{sec:IA_results}

\begin{figure}
    \centering
    \includegraphics[width=\columnwidth]{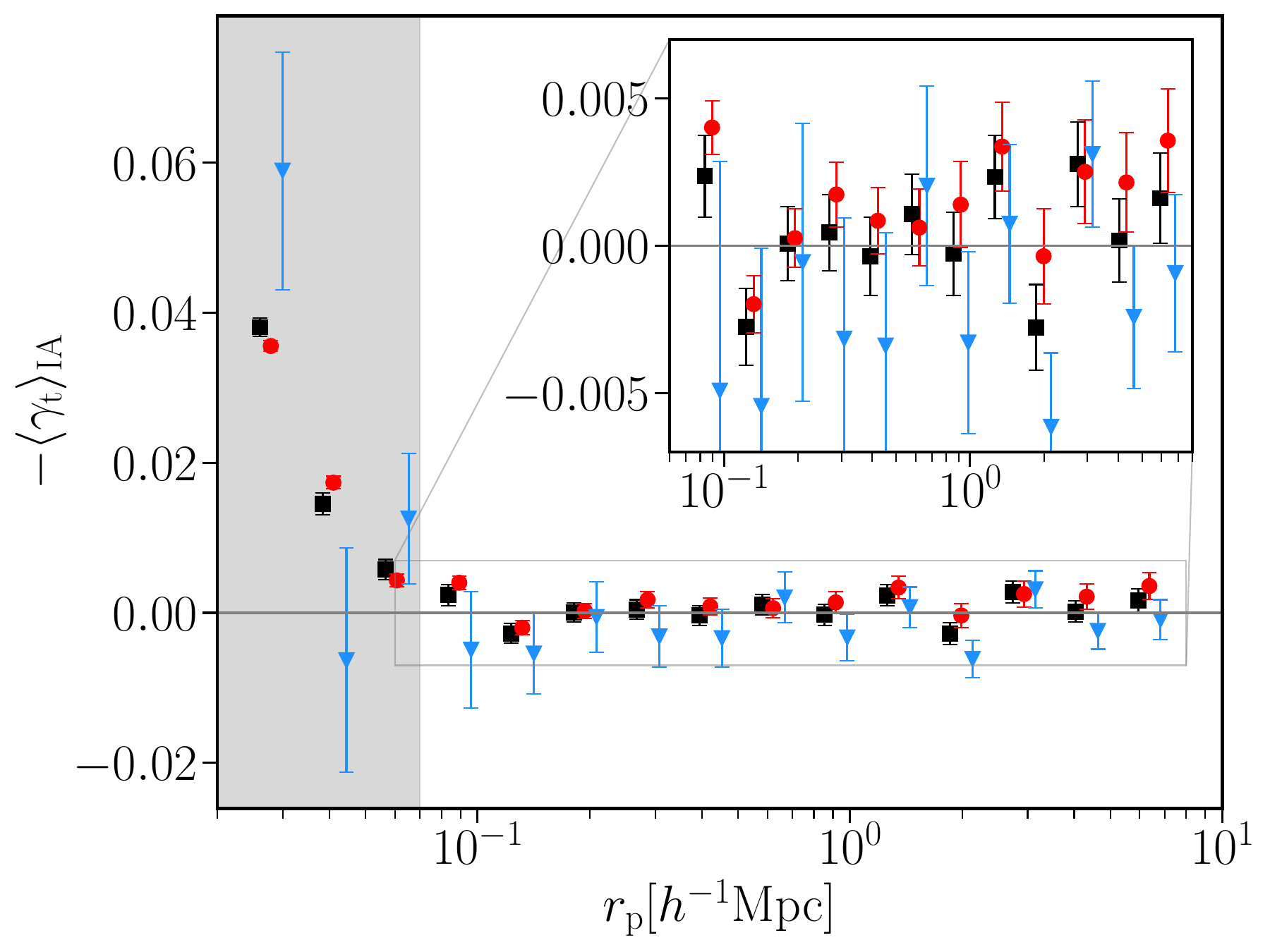}
    \caption{ IA signal of the source galaxy sample around the LRGs, selected in narrow bins in photo-$z$ ($\Delta z_{\rm phot} \equiv |\zl-\zs|<0.06$). We consider the signal from the full source sample (black squares), and its split in red/blue (red circles/blue triangles). At separations below $0.06 \ h^{-1} \rm Mpc$ (grey region) blending becomes important and thus we do not consider it in the analysis. For clarity, the red and blue points are shifted horizontally. }
    \label{fig:ia_signal}
\end{figure}

To constrain the IA of the source galaxies (as a measure of possible satellite galaxy alignment), we measured the $\langle \gamma_{\rm t} \rangle$ signal in narrow bins in photo-$z$ ($\Delta z_{\rm phot} \equiv |\zl-\zs| = 0.06$) and subtracted the lensing contribution (as described in Sect.~\ref{subsec:extracting_IA_signal}). We note that with this definition, the galaxy-galaxy separation is a function of redshift. Here, we are primarily limited by the photometric redshift uncertainty; thus, we consider this effective treatment as sufficient.\footnote{We stress that this IA signal differs from the one discussed in the previous section, which was obtained by using the two-point projected correlation function.} Selecting only galaxies with a small separation along the line of sight enhances the IA signal, while removing a substantial part of the lensing contribution. The different choice in the  $z_{\rm phot}$ binning is fully accounted for in the model through $\Sigma_{\rm crit}$, which was computed according to the new galaxy selection. The mean signal, $\langle \gamma_{\rm t} (r_{\rm p}) \rangle$  was calculated as per Eq.~\eqref{eq:ia_estimator} and, thus, this is the average tangential ellipticity, $\epsilon_{\rm t}$, weighted by the lensing efficiency. Hence, the IA signal is expressed in terms of the weighted intrinsic shear instead of the intrinsic ellipticity, as in \citet[][]{Georgiou2019b}. 

Here, we only consider the lens galaxies from the \dense\ sample, without any split in luminosity. In Appendix~\ref{A:fraction_of_associated_galaxies}, we report the fraction of galaxies of the source sample that are physically associated with the LRGs. The results are presented in Fig.~\ref{fig:ia_signal}.  We do not detect any intrinsic alignment for the full source sample, $\langle \gamma_{t, \rm IA} \rangle = -0.0004 \pm 0.0004$, up to $10 \mpch$. %
As a reference, \citet[][]{Georgiou2019b} found $\langle \epsilon_+ \rangle = 0.0023 \pm 0.0011$\footnote{Average tangential ellipticity. Note:\ the sign convention is opposite in their case, with a positive signal indicating a radial alignment.} (Georgiou, private communication) for the full satellite sample in the Galaxy and Mass Assembly survey \citep[GAMA,][]{Driver2009}. We note that, in their case, the average was computed on the measured $\epsilon_+$ of each satellite galaxy with respect to its central galaxy, for which they used the brightest galaxy of the group as a proxy.
As for the GGL signal, in our average, we did not consider the data points below $0.06 \ \mpch$, where observational systematics become important. 

We go on to explore the possibility that the signal is washed out by the presence of blue galaxies in the sample, which are expected to be poorly or not aligned. We thus split the source galaxies based on their morphology. We followed \citet{Li2021} and used the parameter $T_{\rm B}$ from the BPZ photo-$z$ code \citep[][]{Benitez2000, Coe2006} as a proxy for the morphology. We identified the `red' population (a combination of E1, Sbc, Scd types) as the galaxies satisfying $T_{\rm B} \leq 3$, while the `blue' one is the complementary sample, selected by requiring $T_{\rm B} > 3$. The signal is displayed in Fig.~\ref{fig:ia_signal}. Also, in this case, we did not observe any alignment signal, with $\langle \gamma^{\rm (red)}_{t, \rm IA} \rangle = -0.0015 \pm 0.0004$
and $\langle \gamma^{\rm (blue)}_{t, \rm IA} \rangle = 0.002 \pm 0.001$ (\citet{Georgiou2019b}: $\langle \epsilon_{+, \rm red} \rangle = 0.007$, $\langle \epsilon_{+, \rm blue} \rangle = -0.001$).

To measure the IA with the split in colour, we subtracted the same lensing signal as for the full sample. This is motivated by the fact that lensing does not depend on the source galaxy colour.  However, we tested to see whether the residual IA that could be contaminating the signal would affect the results by measuring the lensing signal only for the red source galaxies; we then measured the IA by subtracting this `red lensing' ESD from $\esd_{\Delta z}^{\rm (red)}$. We found the results to be compatible within the statistical uncertainty with our fiducial setup.

Our results are in line with previous measurements \citep[][]{Hirata2004, Blazek2012, Chisari2014, Sifon2015}. Dark matter-only simulations predict satellite galaxies to be radially aligned towards the centre of the halo \citep[][]{Pereira2008, Pereira2010}, but this signal is significantly washed out when considering the stellar component \citep[][]{Velliscig2015}; this is a plausible consequence of the misalignment between the luminous and the dark component of the galaxy \citep[][]{Velliscig2015a}, even though the observations of halo-galaxy misalignment show no such trend for central galaxies, when considering large enough scales for the luminous matter \citep{Schrabback2015, vanUitert2017, vanUitert2017b, Georgiou2021, Robison2023}. \citet[][]{Pereira&Kuhn2005} and \citet{Faltenbacher2007} both detected a radial alignment signal, the first in cluster galaxies and the second on red satellites in the SDSS galaxy group catalogue \citep[][]{Yang2007}. In line with these findings, recently, \citet[][]{Georgiou2019b} detected a radial signal for both red and blue satellites in galaxy groups selected from the Galaxy And Mass Assembly survey \citep[GAMA,][]{Driver2009}. However, all of these studies rely on spectroscopic redshifts and, thus, on a more robust assignment of the galaxies to their group: here, we considered all galaxies within a given redshift separation, which significantly degrades the signal. Using the photometric redshifts with a cluster finding algorithm can improve such measurements, as shown by \citet{Huang2018} and \citet{Zhou2023}. 

The uncertainty with respect to photometric redshifts also contributes to the dilution of the signal due to the promotion of uncorrelated pairs within the selection, as well as the removal of physically associated galaxies. Moreover, \citet[][]{Velliscig2015a} found that IA depends on the subset of stars used for the signal: using all the stars bound in sub-halos, the signal is significantly increased, compared to the alignment of stars within the half-mass radius. In this latter case, they found compatible values for $\langle \epsilon_{\rm g+}\rangle$ to \citet{Chisari2014, Sifon2015, Singh2015}. This is in line with the finding in \citet[][]{Georgiou2019b}, namely, that the alignment signal is a function of galaxy scale, with the outskirts of the galaxy being more aligned with the position of the central galaxy. In this sense, \lensfit, which gives more weight to the inner part of the galaxy, might also contribute to the low signal observed in our samples. The lack of any IA signal of the source galaxies tells us two things. Firstly, we see that non-LRG galaxies may show a very low IA signal and we are strongly limited by photometric redshifts. Secondly, if we assume that the signal is truly low, this can either be due to the fact that these are less massive and would have much lower alignment amplitude, as shown in Fig. \ref{fig:IA_Meff}; alternatively, they could be actual satellite galaxies that are poorly aligned, with signal that is a strong function of galaxy separation \citep[][]{Pereira2008, Pereira2010}. We compared our results with measurements from GAMA (Georgiou, private communication), but the galaxy selection in that study is different, as those galaxies are truly identified as satellites; whereas in our case, we are accounting for all galaxies in a broad slice of photometric redshifts.

\section{Conclusions}\label{sec:conclusions}

We used weak gravitational lensing to measure the mass of a sample of luminous red galaxies (LRGs) for which the intrinsic alignment (IA) signal was previously measured in \citet{Fortuna2021b}. We split the sample into bins based on their luminosity and used a halo model to interpret our data. We fit the excess surface mass density (ESD) measurements of all the luminosity bins jointly with a single model. 
We confirm that the LRG sample consists mainly of central galaxies, as expected for this kind of galaxy population, and provide the satellite fraction for each $L-$bin. 
We ensured that the modelling recovers the true galaxy properties sufficiently well, by inspecting the luminosity distribution per each luminosity bin predicted by the model. Finally, we find good agreement between the predicted and the real distributions.

The best fit model predicts an increasing average halo mass with luminosity, which we modelled with a double power law and obtained the following slopes: $\gamma_1 = 4.22_{-1.863}^{+1.888}$ and $\gamma_2 = 0.42_{-0.140}^{+0.174}$ (Eq.~\ref{eq:Lc(M)}). We note, however, that our data mainly constrain the high mass-end of the double power law ($M>M_1$), which is reflected by the uncertainties in $\gamma_1$. Our results are in good agreement with previous studies at high luminosity and extend the luminosity-to-halo mass relation towards the faint-end (at low mass).

We used these results to interpret the IA dependence with halo mass, starting from the luminosity dependence measured in the literature. The IA-halo mass relation can be parametrised by a single power law, as predicted by current models \citep[][]{Piras2018}. This suggests that the flattening at low luminosity, hinted at by \citet{Fortuna2021b}, may be caused by the double power law in the luminosity-to-halo mass relation. Although the scatter in the measurements remains large, this would imply that the halo mass is the driving source of the alignment.

We also measured the IA signal of the lensing source sample around the LRGs, by selecting only pairs with a maximum separation of $|\zl-\zs|=0.06$. We considered three cases: the full source sample, and a split in red and blue. We did not detect any alignment in the signal in any of the cases considered for $\rp > 0.06 \mpch$. We mainly attribute our null-detection to the photometric redshift selection of the galaxy pair: the use of better lenses and source photo-$z$ might be useful when revisiting our conclusions.

\begin{acknowledgements}

We thank the anonymous referee for the helpful comments and constructive remarks on this manuscript. MCF and HHo acknowledge support from Vici grant 639.043.512, financed by the Netherlands Organisation for Scientific Research (NWO). AD acknowledges support from the European Research Council under grant agreement No. 770935. HHo also acknowledges funding from the EU Horizon 2020 research and innovation programme under grant agreement 776247. AD and HHi are supported by the ERC Consolidator Grant (No. 770935). HHi is further supported by a DFG Heisenberg grant (Hi 1495/5-1), the DFG Collaborative Research Center SFB1491. NEC is supported through of the project "A rising tide: Galaxy intrinsic alignments as a new probe of cosmology and galaxy evolution'' (with project number VI.Vidi.203.011) of the Talent programme Vidi which is (partly) financed by the Dutch Research Council (NWO). For the purpose of open access, a CC BY public copyright license is applied to any Author Accepted Manuscript version arising from this submission. MA is supported by the UK Science and Technology Facilities Council (STFC) under grant number ST/Y002652/1 and the Royal Society under grant numbers RGSR2222268 and ICAR1231094. MB is supported by the Polish National Science Center through grants no. 2020/38/E/ST9/00395, 2018/30/E/ST9/00698, 2018/31/G/ST9/03388 and 2020/39/B/ST9/03494, and by the Polish Ministry of Science and Higher Education through grant DIR/WK/2018/12. CH acknowledges support from the European Research Council under grant number 647112, from the Max Planck Society and the Alexander von Humboldt Foundation in the framework of the Max Planck-Humboldt Research Award endowed by the Federal Ministry of Education and Research, and the UK Science and Technology Facilities Council (STFC) under grant ST/V000594/1. KK acknowledges support from the Royal Society and Imperial College. AHW is supported by the Deutsches Zentrum für Luft- und Raumfahrt (DLR), made possible by the Bundesministerium für Wirtschaft und Klimaschutz, and acknowledges funding from the German Science Foundation DFG, via the Collaborative Research Center SFB1491 “Cosmic Interacting Matters - From Source to Signal”. JY acknowledges the support of the National Science Foundation of China (12203084), the China Postdoctoral Science Foundation (2021T140451), and the Shanghai Post-doctoral Excellence Program (2021419). \\

Based on data products from observations made with ESO Telescopes at the La Silla Paranal Observatory under programme IDs 177.A- 3016, 177.A-3017 and 177.A-3018, and on data products produced by Tar- get/OmegaCEN, INAF-OACN, INAF-OAPD and the KiDS production team, on behalf of the KiDS consortium. OmegaCEN and the KiDS production team acknowledge support by NOVA and NWO-M grants. Members of INAF-OAPD and INAF-OACN also acknowledge the support from the Department of Physics $\&$ Astronomy of the University of Padova, and of the Department of Physics of Univ. Federico II (Naples). \\

\textit{Author contributions:} All authors contributed to the development and writing of this paper. The authorship list is given in three groups: the lead authors (MCF, AD, HH) followed by two alphabetical groups. The first alphabetical group includes those who are key contributors to both the scientific analysis and the data products. The second group covers those who have either made a significant contribution to the data products, or to the scientific analysis.

\end{acknowledgements}



\bibliographystyle{aa}
\bibliography{gglbiblio}

\begin{thebibliography}{102}
\expandafter\ifx\csname natexlab\endcsname\relax\def\natexlab#1{#1}\fi

\bibitem[{{Arnouts} {et~al.}(1999){Arnouts}, {Cristiani}, {Moscardini},
  {Matarrese}, {Lucchin}, {Fontana}, \& {Giallongo}}]{Arnouts1999}
{Arnouts}, S., {Cristiani}, S., {Moscardini}, L., {et~al.} 1999, \mnras, 310,
  540

\bibitem[{{Arnouts} \& {Ilbert}(2011)}]{Arnouts2011Lephare}
{Arnouts}, S. \& {Ilbert}, O. 2011, {LePHARE: Photometric Analysis for Redshift
  Estimate}

\bibitem[{{Asgari} {et~al.}(2023){Asgari}, {Mead}, \&
  {Heymans}}]{2023Asgari_HaloModelReview}
{Asgari}, M., {Mead}, A.~J., \& {Heymans}, C. 2023, arXiv e-prints,
  arXiv:2303.08752

\bibitem[{{Ben{\'\i}tez}(2000)}]{Benitez2000}
{Ben{\'\i}tez}, N. 2000, \apj, 536, 571

\bibitem[{{Benson} {et~al.}(2000){Benson}, {Cole}, {Frenk}, {Baugh}, \&
  {Lacey}}]{Benson2000}
{Benson}, A.~J., {Cole}, S., {Frenk}, C.~S., {Baugh}, C.~M., \& {Lacey}, C.~G.
  2000, \mnras, 311, 793

\bibitem[{{Berlind} \& {Weinberg}(2002)}]{Berlind2002}
{Berlind}, A.~A. \& {Weinberg}, D.~H. 2002, \apj, 575, 587

\bibitem[{{Bilicki} {et~al.}(2021){Bilicki}, {Dvornik}, {Hoekstra}, {Wright},
  {Chisari}, {Vakili}, {Asgari}, {Giblin}, {Heymans}, {Hildebrandt},
  {Holwerda}, {Hopkins}, {Johnston}, {Kannawadi}, {Kuijken}, {Nakoneczny},
  {Shan}, {Sonnenfeld}, \& {Valentijn}}]{Bilicki2021}
{Bilicki}, M., {Dvornik}, A., {Hoekstra}, H., {et~al.} 2021, \aap, 653, A82

\bibitem[{{Blazek} {et~al.}(2012){Blazek}, {Mandelbaum}, {Seljak}, \&
  {Nakajima}}]{Blazek2012}
{Blazek}, J., {Mandelbaum}, R., {Seljak}, U., \& {Nakajima}, R. 2012, \jcap,
  2012, 041

\bibitem[{{Blazek} {et~al.}(2011){Blazek}, {McQuinn}, \& {Seljak}}]{Blazek2011}
{Blazek}, J., {McQuinn}, M., \& {Seljak}, U. 2011, \jcap, 2011, 010

\bibitem[{Blazek {et~al.}(2019)Blazek, MacCrann, Troxel, \& Fang}]{Blazek2017}
Blazek, J.~A., MacCrann, N., Troxel, M.~A., \& Fang, X. 2019, Phys. Rev. D,
  100, 103506

\bibitem[{{Bridle} \& {King}(2007)}]{BridleKing2007}
{Bridle}, S. \& {King}, L. 2007, New Journal of Physics, 9, 444

\bibitem[{{Bruzual} \& {Charlot}(2003)}]{BruzualCharlot2003}
{Bruzual}, G. \& {Charlot}, S. 2003, \mnras, 344, 1000

\bibitem[{{Cacciato} {et~al.}(2009){Cacciato}, {van den Bosch}, {More}, {Li},
  {Mo}, \& {Yang}}]{Cacciato2009}
{Cacciato}, M., {van den Bosch}, F.~C., {More}, S., {et~al.} 2009, \mnras, 394,
  929

\bibitem[{{Cacciato} {et~al.}(2013){Cacciato}, {van den Bosch}, {More}, {Mo},
  \& {Yang}}]{Cacciato2013}
{Cacciato}, M., {van den Bosch}, F.~C., {More}, S., {Mo}, H., \& {Yang}, X.
  2013, \mnras, 430, 767

\bibitem[{{Cacciato} {et~al.}(2014){Cacciato}, {van Uitert}, \&
  {Hoekstra}}]{Cacciato2014}
{Cacciato}, M., {van Uitert}, E., \& {Hoekstra}, H. 2014, \mnras, 437, 377

\bibitem[{{Calzetti} {et~al.}(1994){Calzetti}, {Kinney}, \&
  {Storchi-Bergmann}}]{Calzetti1994}
{Calzetti}, D., {Kinney}, A.~L., \& {Storchi-Bergmann}, T. 1994, \apj, 429, 582

\bibitem[{{Capaccioli} {et~al.}(2012){Capaccioli}, {Schipani}, {de Paris},
  {Grado}, {Napolitano}, {Iodice}, {Marconi}, {Merluzzi}, \&
  {Ripepi}}]{Capaccioli2012}
{Capaccioli}, M., {Schipani}, P., {de Paris}, G., {et~al.} 2012, in Science
  from the Next Generation Imaging and Spectroscopic Surveys, 1

\bibitem[{Carretero {et~al.}(2017)}]{Carretero2017CosmoHub}
Carretero, J. {et~al.} 2017, PoS, EPS-HEP2017, 488

\bibitem[{{Catelan} {et~al.}(2001){Catelan}, {Kamionkowski}, \&
  {Blandford}}]{Catelan2001}
{Catelan}, P., {Kamionkowski}, M., \& {Blandford}, R.~D. 2001, \mnras, 320, L7

\bibitem[{{Chabrier}(2003)}]{Chabrier2003}
{Chabrier}, G. 2003, \pasp, 115, 763

\bibitem[{{Chisari} {et~al.}(2014){Chisari}, {Mandelbaum}, {Strauss}, {Huff},
  \& {Bahcall}}]{Chisari2014}
{Chisari}, N.~E., {Mandelbaum}, R., {Strauss}, M.~A., {Huff}, E.~M., \&
  {Bahcall}, N.~A. 2014, \mnras, 445, 726

\bibitem[{{Coe} {et~al.}(2006){Coe}, {Ben{\'\i}tez}, {S{\'a}nchez}, {Jee},
  {Bouwens}, \& {Ford}}]{Coe2006}
{Coe}, D., {Ben{\'\i}tez}, N., {S{\'a}nchez}, S.~F., {et~al.} 2006, \aj, 132,
  926

\bibitem[{Cooray \& Sheth(2002)}]{CooraySheth2002}
Cooray, A. \& Sheth, R. 2002, Phys. Rep., 372, 1

\bibitem[{{Croft} \& {Metzler}(2000)}]{Croft&Metzer2000}
{Croft}, R. A.~C. \& {Metzler}, C.~A. 2000, \apj, 545, 561

\bibitem[{{Driver} {et~al.}(2009){Driver}, {Norberg}, {Baldry}, {Bamford},
  {Hopkins}, {Liske}, {Loveday}, {Peacock}, {Hill}, {Kelvin}, {Robotham},
  {Cross}, {Parkinson}, {Prescott}, {Conselice}, {Dunne}, {Brough}, {Jones},
  {Sharp}, {van Kampen}, {Oliver}, {Roseboom}, {Bland-Hawthorn}, {Croom},
  {Ellis}, {Cameron}, {Cole}, {Frenk}, {Couch}, {Graham}, {Proctor}, {De
  Propris}, {Doyle}, {Edmondson}, {Nichol}, {Thomas}, {Eales}, {Jarvis},
  {Kuijken}, {Lahav}, {Madore}, {Seibert}, {Meyer}, {Staveley-Smith},
  {Phillipps}, {Popescu}, {Sansom}, {Sutherland}, {Tuffs}, \&
  {Warren}}]{Driver2009}
{Driver}, S.~P., {Norberg}, P., {Baldry}, I.~K., {et~al.} 2009, Astronomy and
  Geophysics, 50, 5.12

\bibitem[{{Dubinski}(1992)}]{Dubinski1992}
{Dubinski}, J. 1992, \apj, 401, 441

\bibitem[{{Duffy} {et~al.}(2008){Duffy}, {Schaye}, {Kay}, \& {Dalla
  Vecchia}}]{Duffy2008}
{Duffy}, A.~R., {Schaye}, J., {Kay}, S.~T., \& {Dalla Vecchia}, C. 2008,
  \mnras, 390, L64

\bibitem[{{Dvornik} {et~al.}(2018){Dvornik}, {Hoekstra}, {Kuijken},
  {Schneider}, {Amon}, {Nakajima}, {Viola}, {Choi}, {Erben}, {Farrow},
  {Heymans}, {Hildebrandt}, {Sif{\'o}n}, \& {Wang}}]{Dvornik2018}
{Dvornik}, A., {Hoekstra}, H., {Kuijken}, K., {et~al.} 2018, \mnras, 479, 1240

\bibitem[{{Edge} {et~al.}(2013){Edge}, {Sutherland}, {Kuijken}, {Driver},
  {McMahon}, {Eales}, \& {Emerson}}]{Edge2013}
{Edge}, A., {Sutherland}, W., {Kuijken}, K., {et~al.} 2013, The Messenger, 154,
  32

\bibitem[{{Faltenbacher} {et~al.}(2007){Faltenbacher}, {Li}, {Mao}, {van den
  Bosch}, {Yang}, {Jing}, {Pasquali}, \& {Mo}}]{Faltenbacher2007}
{Faltenbacher}, A., {Li}, C., {Mao}, S., {et~al.} 2007, \apjl, 662, L71

\bibitem[{{Faltenbacher} {et~al.}(2009){Faltenbacher}, {Li}, {White}, {Jing},
  {Shu-DeMao}, \& {Wang}}]{Faltenbacher2009}
{Faltenbacher}, A., {Li}, C., {White}, S. D.~M., {et~al.} 2009, Research in
  Astronomy and Astrophysics, 9, 41

\bibitem[{{Fenech Conti} {et~al.}(2017){Fenech Conti}, {Herbonnet}, {Hoekstra},
  {Merten}, {Miller}, \& {Viola}}]{FenechConti2017}
{Fenech Conti}, I., {Herbonnet}, R., {Hoekstra}, H., {et~al.} 2017, \mnras,
  467, 1627

\bibitem[{{Foreman-Mackey} {et~al.}(2013){Foreman-Mackey}, {Hogg}, {Lang}, \&
  {Goodman}}]{Foreman-Mackey2013emcee}
{Foreman-Mackey}, D., {Hogg}, D.~W., {Lang}, D., \& {Goodman}, J. 2013, \pasp,
  125, 306

\bibitem[{{Forero-Romero} {et~al.}(2014){Forero-Romero}, {Contreras}, \&
  {Padilla}}]{Forero-Romero2014}
{Forero-Romero}, J.~E., {Contreras}, S., \& {Padilla}, N. 2014, \mnras, 443,
  1090

\bibitem[{{Fortuna} {et~al.}(2020){Fortuna}, {Hoekstra}, {Joachimi},
  {Johnston}, {Chisari}, {Georgiou}, \& {Mahony}}]{fortuna2020halo}
{Fortuna}, M.~C., {Hoekstra}, H., {Joachimi}, B., {et~al.} 2020, \mnras
  [\eprint[arXiv]{2003.02700}]

\bibitem[{{Fortuna} {et~al.}(2021){Fortuna}, {Hoekstra}, {Johnston}, {Vakili},
  {Kannawadi}, {Georgiou}, {Joachimi}, {Wright}, {Asgari}, {Bilicki},
  {Heymans}, {Hildebrandt}, {Kuijken}, \& {Von
  Wietersheim-Kramsta}}]{Fortuna2021b}
{Fortuna}, M.~C., {Hoekstra}, H., {Johnston}, H., {et~al.} 2021, \aap, 654, A76

\bibitem[{{Gaia Collaboration} {et~al.}(2018){Gaia Collaboration}, {Brown},
  {Vallenari}, {Prusti}, {de Bruijne}, {Babusiaux}, {Bailer-Jones}, {Biermann},
  {Evans}, {Eyer}, {Jansen}, {Jordi}, {Klioner}, {Lammers}, {Lindegren},
  {Luri}, {Mignard}, {Panem}, {Pourbaix}, {Randich}, {Sartoretti}, {Siddiqui},
  {Soubiran}, {van Leeuwen}, {Walton}, {Arenou}, {Bastian}, {Cropper},
  {Drimmel}, {Katz}, {Lattanzi}, {Bakker}, {Cacciari}, {Casta{\~n}eda},
  {Chaoul}, {Cheek}, {De Angeli}, {Fabricius}, {Guerra}, {Holl}, {Masana},
  {Messineo}, {Mowlavi}, {Nienartowicz}, {Panuzzo}, {Portell}, {Riello},
  {Seabroke}, {Tanga}, {Th{\'e}venin}, {Gracia-Abril}, {Comoretto},
  {Garcia-Reinaldos}, {Teyssier}, {Altmann}, {Andrae}, {Audard},
  {Bellas-Velidis}, {Benson}, {Berthier}, {Blomme}, {Burgess}, {Busso},
  {Carry}, {Cellino}, {Clementini}, {Clotet}, {Creevey}, {Davidson}, {De
  Ridder}, {Delchambre}, {Dell'Oro}, {Ducourant},
  {Fern{\'a}ndez-Hern{\'a}ndez}, {Fouesneau}, {Fr{\'e}mat}, {Galluccio},
  {Garc{\'\i}a-Torres}, {Gonz{\'a}lez-N{\'u}{\~n}ez}, {Gonz{\'a}lez-Vidal},
  {Gosset}, {Guy}, {Halbwachs}, {Hambly}, {Harrison}, {Hern{\'a}ndez},
  {Hestroffer}, {Hodgkin}, {Hutton}, {Jasniewicz}, {Jean-Antoine-Piccolo},
  {Jordan}, {Korn}, {Krone-Martins}, {Lanzafame}, {Lebzelter}, {L{\"o}ffler},
  {Manteiga}, {Marrese}, {Mart{\'\i}n-Fleitas}, {Moitinho}, {Mora}, {Muinonen},
  {Osinde}, {Pancino}, {Pauwels}, {Petit}, {Recio-Blanco}, {Richards},
  {Rimoldini}, {Robin}, {Sarro}, {Siopis}, {Smith}, {Sozzetti}, {S{\"u}veges},
  {Torra}, {van Reeven}, {Abbas}, {Abreu Aramburu}, {Accart}, {Aerts},
  {Altavilla}, {{\'A}lvarez}, {Alvarez}, {Alves}, {Anderson}, {Andrei},
  {Anglada Varela}, {Antiche}, {Antoja}, {Arcay}, {Astraatmadja}, {Bach},
  {Baker}, {Balaguer-N{\'u}{\~n}ez}, {Balm}, {Barache}, {Barata}, {Barbato},
  {Barblan}, {Barklem}, {Barrado}, {Barros}, {Barstow}, {Bartholom{\'e}
  Mu{\~n}oz}, {Bassilana}, {Becciani}, {Bellazzini}, {Berihuete}, {Bertone},
  {Bianchi}, {Bienaym{\'e}}, {Blanco-Cuaresma}, {Boch}, {Boeche}, {Bombrun},
  {Borrachero}, {Bossini}, {Bouquillon}, {Bourda}, {Bragaglia}, {Bramante},
  {Breddels}, {Bressan}, {Brouillet}, {Br{\"u}semeister}, {Brugaletta},
  {Bucciarelli}, {Burlacu}, {Busonero}, {Butkevich}, {Buzzi}, {Caffau},
  {Cancelliere}, {Cannizzaro}, {Cantat-Gaudin}, {Carballo}, {Carlucci},
  {Carrasco}, {Casamiquela}, {Castellani}, {Castro-Ginard}, {Charlot},
  {Chemin}, {Chiavassa}, {Cocozza}, {Costigan}, {Cowell}, {Crifo}, {Crosta},
  {Crowley}, {Cuypers}, {Dafonte}, {Damerdji}, {Dapergolas}, {David}, {David},
  {de Laverny}, {De Luise}, {De March}, {de Martino}, {de Souza}, {de Torres},
  {Debosscher}, {del Pozo}, {Delbo}, {Delgado}, {Delgado}, {Di Matteo},
  {Diakite}, {Diener}, {Distefano}, {Dolding}, {Drazinos}, {Dur{\'a}n},
  {Edvardsson}, {Enke}, {Eriksson}, {Esquej}, {Eynard Bontemps}, {Fabre},
  {Fabrizio}, {Faigler}, {Falc{\~a}o}, {Farr{\`a}s Casas}, {Federici},
  {Fedorets}, {Fernique}, {Figueras}, {Filippi}, {Findeisen}, {Fonti},
  {Fraile}, {Fraser}, {Fr{\'e}zouls}, {Gai}, {Galleti}, {Garabato},
  {Garc{\'\i}a-Sedano}, {Garofalo}, {Garralda}, {Gavel}, {Gavras}, {Gerssen},
  {Geyer}, {Giacobbe}, {Gilmore}, {Girona}, {Giuffrida}, {Glass}, {Gomes},
  {Granvik}, {Gueguen}, {Guerrier}, {Guiraud}, {Guti{\'e}rrez-S{\'a}nchez},
  {Haigron}, {Hatzidimitriou}, {Hauser}, {Haywood}, {Heiter}, {Helmi}, {Heu},
  {Hilger}, {Hobbs}, {Hofmann}, {Holland}, {Huckle}, {Hypki}, {Icardi},
  {Jan{\ss}en}, {Jevardat de Fombelle}, {Jonker}, {Juh{\'a}sz}, {Julbe},
  {Karampelas}, {Kewley}, {Klar}, {Kochoska}, {Kohley}, {Kolenberg},
  {Kontizas}, {Kontizas}, {Koposov}, {Kordopatis}, {Kostrzewa-Rutkowska},
  {Koubsky}, {Lambert}, {Lanza}, {Lasne}, {Lavigne}, {Le Fustec}, {Le
  Poncin-Lafitte}, {Lebreton}, {Leccia}, {Leclerc}, {Lecoeur-Taibi},
  {Lenhardt}, {Leroux}, {Liao}, {Licata}, {Lindstr{\o}m}, {Lister}, {Livanou},
  {Lobel}, {L{\'o}pez}, {Managau}, {Mann}, {Mantelet}, {Marchal}, {Marchant},
  {Marconi}, {Marinoni}, {Marschalk{\'o}}, {Marshall}, {Martino}, {Marton},
  {Mary}, {Massari}, {Matijevi{\v{c}}}, {Mazeh}, {McMillan}, {Messina},
  {Michalik}, {Millar}, {Molina}, {Molinaro}, {Moln{\'a}r}, {Montegriffo},
  {Mor}, {Morbidelli}, {Morel}, {Morris}, {Mulone}, {Muraveva}, {Musella},
  {Nelemans}, {Nicastro}, {Noval}, {O'Mullane}, {Ord{\'e}novic},
  {Ord{\'o}{\~n}ez-Blanco}, {Osborne}, {Pagani}, {Pagano}, {Pailler},
  {Palacin}, {Palaversa}, {Panahi}, {Pawlak}, {Piersimoni}, {Pineau}, {Plachy},
  {Plum}, {Poggio}, {Poujoulet}, {Pr{\v{s}}a}, {Pulone}, {Racero}, {Ragaini},
  {Rambaux}, {Ramos-Lerate}, {Regibo}, {Reyl{\'e}}, {Riclet}, {Ripepi}, {Riva},
  {Rivard}, {Rixon}, {Roegiers}, {Roelens}, {Romero-G{\'o}mez}, {Rowell},
  {Royer}, {Ruiz-Dern}, {Sadowski}, {Sagrist{\`a} Sell{\'e}s}, {Sahlmann},
  {Salgado}, {Salguero}, {Sanna}, {Santana-Ros}, {Sarasso}, {Savietto},
  {Schultheis}, {Sciacca}, {Segol}, {Segovia}, {S{\'e}gransan}, {Shih},
  {Siltala}, {Silva}, {Smart}, {Smith}, {Solano}, {Solitro}, {Sordo}, {Soria
  Nieto}, {Souchay}, {Spagna}, {Spoto}, {Stampa}, {Steele},
  {Steidelm{\"u}ller}, {Stephenson}, {Stoev}, {Suess}, {Surdej}, {Szabados},
  {Szegedi-Elek}, {Tapiador}, {Taris}, {Tauran}, {Taylor}, {Teixeira},
  {Terrett}, {Teyssandier}, {Thuillot}, {Titarenko}, {Torra Clotet}, {Turon},
  {Ulla}, {Utrilla}, {Uzzi}, {Vaillant}, {Valentini}, {Valette}, {van Elteren},
  {Van Hemelryck}, {van Leeuwen}, {Vaschetto}, {Vecchiato}, {Veljanoski},
  {Viala}, {Vicente}, {Vogt}, {von Essen}, {Voss}, {Votruba}, {Voutsinas},
  {Walmsley}, {Weiler}, {Wertz}, {Wevers}, {Wyrzykowski}, {Yoldas},
  {{\v{Z}}erjal}, {Ziaeepour}, {Zorec}, {Zschocke}, {Zucker}, {Zurbach}, \&
  {Zwitter}}]{Brown2018GaiaDR2}
{Gaia Collaboration}, {Brown}, A.~G.~A., {Vallenari}, A., {et~al.} 2018, \aap,
  616, A1

\bibitem[{{Georgiou} {et~al.}(2019){Georgiou}, {Chisari}, {Fortuna},
  {Hoekstra}, {Kuijken}, {Joachimi}, {Vakili}, {Bilicki}, {Dvornik}, {Erben},
  {Giblin}, {Heymans}, {Napolitano}, \& {Shan}}]{Georgiou2019b}
{Georgiou}, C., {Chisari}, N.~E., {Fortuna}, M.~C., {et~al.} 2019, \aap, 628,
  A31

\bibitem[{{Georgiou} {et~al.}(2021){Georgiou}, {Hoekstra}, {Kuijken},
  {Bilicki}, {Dvornik}, {Erben}, {Giblin}, {Heymans}, {Hildebrandt}, {de Jong},
  {Kannawadi}, {Schneider}, {Schrabback}, {Shan}, \& {Wright}}]{Georgiou2021}
{Georgiou}, C., {Hoekstra}, H., {Kuijken}, K., {et~al.} 2021, \aap, 647, A185

\bibitem[{{Giblin} {et~al.}(2021){Giblin}, {Heymans}, {Asgari}, {Hildebrandt},
  {Hoekstra}, {Joachimi}, {Kannawadi}, {Kuijken}, {Lin}, {Miller},
  {Tr{\"o}ster}, {van den Busch}, {Wright}, {Bilicki}, {Blake}, {de Jong},
  {Dvornik}, {Erben}, {Getman}, {Napolitano}, {Schneider}, {Shan}, \&
  {Valentijn}}]{Giblin2021}
{Giblin}, B., {Heymans}, C., {Asgari}, M., {et~al.} 2021, \aap, 645, A105

\bibitem[{{Hahn} {et~al.}(2007){Hahn}, {Porciani}, {Carollo}, \&
  {Dekel}}]{Oliver&Porciani2007}
{Hahn}, O., {Porciani}, C., {Carollo}, C.~M., \& {Dekel}, A. 2007, \mnras, 375,
  489

\bibitem[{{Hildebrandt} {et~al.}(2020){Hildebrandt}, {K{\"o}hlinger}, {van den
  Busch}, {Joachimi}, {Heymans}, {Kannawadi}, {Wright}, {Asgari}, {Blake},
  {Hoekstra}, {Joudaki}, {Kuijken}, {Miller}, {Morrison}, {Tr{\"o}ster},
  {Amon}, {Archidiacono}, {Brieden}, {Choi}, {de Jong}, {Erben}, {Giblin},
  {Mead}, {Peacock}, {Radovich}, {Schneider}, {Sif{\'o}n}, \&
  {Tewes}}]{Hildebrandt2020}
{Hildebrandt}, H., {K{\"o}hlinger}, F., {van den Busch}, J.~L., {et~al.} 2020,
  \aap, 633, A69

\bibitem[{{Hildebrandt} {et~al.}(2021){Hildebrandt}, {van den Busch}, {Wright},
  {Blake}, {Joachimi}, {Kuijken}, {Tr{\"o}ster}, {Asgari}, {Bilicki}, {de
  Jong}, {Dvornik}, {Erben}, {Getman}, {Giblin}, {Heymans}, {Kannawadi}, {Lin},
  \& {Shan}}]{Hildebrandt2021photoz}
{Hildebrandt}, H., {van den Busch}, J.~L., {Wright}, A.~H., {et~al.} 2021,
  \aap, 647, A124

\bibitem[{{Hirata} {et~al.}(2007){Hirata}, {Mandelbaum}, {Ishak}, {Seljak},
  {Nichol}, {Pimbblet}, {Ross}, \& {Wake}}]{Hirata2007}
{Hirata}, C.~M., {Mandelbaum}, R., {Ishak}, M., {et~al.} 2007, \mnras, 381,
  1197

\bibitem[{{Hirata} \& {Seljak}(2004)}]{Hirata2004}
{Hirata}, C.~M. \& {Seljak}, U. 2004, \prd, 70, 063526

\bibitem[{{Hoekstra} {et~al.}(2005){Hoekstra}, {Hsieh}, {Yee}, {Lin}, \&
  {Gladders}}]{Hoekstra2005}
{Hoekstra}, H., {Hsieh}, B.~C., {Yee}, H.~K.~C., {Lin}, H., \& {Gladders},
  M.~D. 2005, \apj, 635, 73

\bibitem[{{Hopkins} {et~al.}(2005){Hopkins}, {Bahcall}, \&
  {Bode}}]{Hopkins2005}
{Hopkins}, P.~F., {Bahcall}, N.~A., \& {Bode}, P. 2005, \apj, 618, 1

\bibitem[{{Huang} {et~al.}(2018){Huang}, {Mandelbaum}, {Freeman}, {Chen},
  {Rozo}, \& {Rykoff}}]{Huang2018}
{Huang}, H.-J., {Mandelbaum}, R., {Freeman}, P.~E., {et~al.} 2018, \mnras, 474,
  4772

\bibitem[{{Ilbert} {et~al.}(2006){Ilbert}, {Arnouts}, {McCracken},
  {Bolzonella}, {Bertin}, {Le F{\`e}vre}, {Mellier}, {Zamorani}, {Pell{\`o}},
  {Iovino}, {Tresse}, {Le Brun}, {Bottini}, {Garilli}, {Maccagni}, {Picat},
  {Scaramella}, {Scodeggio}, {Vettolani}, {Zanichelli}, {Adami}, {Bardelli},
  {Cappi}, {Charlot}, {Ciliegi}, {Contini}, {Cucciati}, {Foucaud}, {Franzetti},
  {Gavignaud}, {Guzzo}, {Marano}, {Marinoni}, {Mazure}, {Meneux}, {Merighi},
  {Paltani}, {Pollo}, {Pozzetti}, {Radovich}, {Zucca}, {Bondi}, {Bongiorno},
  {Busarello}, {de La Torre}, {Gregorini}, {Lamareille}, {Mathez}, {Merluzzi},
  {Ripepi}, {Rizzo}, \& {Vergani}}]{Ilbert2006}
{Ilbert}, O., {Arnouts}, S., {McCracken}, H.~J., {et~al.} 2006, \aap, 457, 841

\bibitem[{Joachimi {et~al.}(2011)Joachimi, Mandelbaum, Abdalla, \&
  Bridle}]{Joachimi2011b}
Joachimi, B., Mandelbaum, R., Abdalla, F.~B., \& Bridle, S.~L. 2011, A{\&}A,
  527

\bibitem[{{Johnston} {et~al.}(2019){Johnston}, {Georgiou}, {Joachimi},
  {Hoekstra}, {Chisari}, {Farrow}, {Fortuna}, {Heymans}, {Joudaki}, {Kuijken},
  \& {Wright}}]{Johnston2019}
{Johnston}, H., {Georgiou}, C., {Joachimi}, B., {et~al.} 2019, \aap, 624, A30

\bibitem[{{Kannawadi} {et~al.}(2019){Kannawadi}, {Hoekstra}, {Miller}, {Viola},
  {Fenech Conti}, {Herbonnet}, {Erben}, {Heymans}, {Hildebrandt}, {Kuijken},
  {Vakili}, \& {Wright}}]{Kannawadi2019}
{Kannawadi}, A., {Hoekstra}, H., {Miller}, L., {et~al.} 2019, \aap, 624, A92

\bibitem[{{Kauffmann} {et~al.}(1999){Kauffmann}, {Colberg}, {Diaferio}, \&
  {White}}]{Kauffmann1999}
{Kauffmann}, G., {Colberg}, J.~M., {Diaferio}, A., \& {White}, S. D.~M. 1999,
  \mnras, 303, 188

\bibitem[{{Kuijken}(2011)}]{Kuijken2011}
{Kuijken}, K. 2011, The Messenger, 146, 8

\bibitem[{{Kuijken} {et~al.}(2019){Kuijken}, {Heymans}, {Dvornik},
  {Hildebrandt}, {de Jong}, {Wright}, {Erben}, {Bilicki}, {Giblin}, {Shan},
  {Getman}, {Grado}, {Hoekstra}, {Miller}, {Napolitano}, {Paolilo}, {Radovich},
  {Schneider}, {Sutherland }, {Tewes}, {Tortora}, {Valentijn}, \& {Verdoes
  Kleijn}}]{Kuijken2019DR4}
{Kuijken}, K., {Heymans}, C., {Dvornik}, A., {et~al.} 2019, \aap, 625, A2

\bibitem[{{Leauthaud} {et~al.}(2017){Leauthaud}, {Saito}, {Hilbert},
  {Barreira}, {More}, {White}, {Alam}, {Behroozi}, {Bundy}, {Coupon}, {Erben},
  {Heymans}, {Hildebrandt}, {Mandelbaum}, {Miller}, {Moraes}, {Pereira},
  {Rodr{\'\i}guez-Torres}, {Schmidt}, {Shan}, {Viel}, \&
  {Villaescusa-Navarro}}]{Leauthaud2017LensingIsLow}
{Leauthaud}, A., {Saito}, S., {Hilbert}, S., {et~al.} 2017, \mnras, 467, 3024

\bibitem[{{Lee} {et~al.}(2008){Lee}, {Springel}, {Pen}, \& {Lemson}}]{Lee2008}
{Lee}, J., {Springel}, V., {Pen}, U.-L., \& {Lemson}, G. 2008, \mnras, 389,
  1266

\bibitem[{{Leonard} {et~al.}(2018){Leonard}, {Mandelbaum}, \& {LSST Dark Energy
  Science Collaboration}}]{Leonard2018}
{Leonard}, C.~D., {Mandelbaum}, R., \& {LSST Dark Energy Science
  Collaboration}. 2018, \mnras, 479, 1412

\bibitem[{{Li} {et~al.}(2021){Li}, {Kuijken}, {Hoekstra}, {Hildebrandt},
  {Joachimi}, \& {Kannawadi}}]{Li2021}
{Li}, S.-S., {Kuijken}, K., {Hoekstra}, H., {et~al.} 2021, \aap, 646, A175

\bibitem[{Ma \& Fry(2000)}]{MaFry2000}
Ma, C.-P. \& Fry, J.~N. 2000, The Astrophysical Journal, 543, 503

\bibitem[{{Mancone} \& {Gonzalez}(2012)}]{Mancone2012EzGal}
{Mancone}, C.~L. \& {Gonzalez}, A.~H. 2012, \pasp, 124, 606

\bibitem[{{Mandelbaum} {et~al.}(2011){Mandelbaum}, {Blake}, {Bridle},
  {Abdalla}, {Brough}, {Colless}, {Couch}, {Croom}, {Davis}, {Drinkwater},
  {Forster}, {Glazebrook}, {Jelliffe}, {Jurek}, {Li}, {Madore}, {Martin},
  {Pimbblet}, {Poole}, {Pracy}, {Sharp}, {Wisnioski}, {Woods}, \&
  {Wyder}}]{Mandelbaum2011WiggleZ}
{Mandelbaum}, R., {Blake}, C., {Bridle}, S., {et~al.} 2011, \mnras, 410, 844

\bibitem[{Mandelbaum {et~al.}(2006)Mandelbaum, Hirata, Ishak, Seljak, \&
  Brinkmann}]{Mandelbaum2006a}
Mandelbaum, R., Hirata, C.~M., Ishak, M., Seljak, U., \& Brinkmann, J. 2006,
  Mon. Not. R. Astron. Soc, 367, 611

\bibitem[{{Mandelbaum} {et~al.}(2005){Mandelbaum}, {Hirata}, {Seljak}, {Guzik},
  {Padmanabhan}, {Blake}, {Blanton}, {Lupton}, \& {Brinkmann}}]{Mandelbaum2005}
{Mandelbaum}, R., {Hirata}, C.~M., {Seljak}, U., {et~al.} 2005, \mnras, 361,
  1287

\bibitem[{{Miller} {et~al.}(2013){Miller}, {Heymans}, {Kitching}, {van
  Waerbeke}, {Erben}, {Hildebrandt}, {Hoekstra}, {Mellier}, {Rowe}, {Coupon},
  {Dietrich}, {Fu}, {Harnois-D{\'e}raps}, {Hudson}, {Kilbinger}, {Kuijken},
  {Schrabback}, {Semboloni}, {Vafaei}, \& {Veland er}}]{Miller2013}
{Miller}, L., {Heymans}, C., {Kitching}, T.~D., {et~al.} 2013, \mnras, 429,
  2858

\bibitem[{{Miller} {et~al.}(2007){Miller}, {Kitching}, {Heymans}, {Heavens}, \&
  {van Waerbeke}}]{Miller2007}
{Miller}, L., {Kitching}, T.~D., {Heymans}, C., {Heavens}, A.~F., \& {van
  Waerbeke}, L. 2007, \mnras, 382, 315

\bibitem[{{Miyatake} {et~al.}(2015){Miyatake}, {More}, {Mandelbaum}, {Takada},
  {Spergel}, {Kneib}, {Schneider}, {Brinkmann}, \& {Brownstein}}]{Miyatake2015}
{Miyatake}, H., {More}, S., {Mandelbaum}, R., {et~al.} 2015, \apj, 806, 1

\bibitem[{{Navarro} {et~al.}(1996){Navarro}, {Frenk}, \& {White}}]{NFW1996}
{Navarro}, J.~F., {Frenk}, C.~S., \& {White}, S.~D.~M. 1996, \apj, 462, 563

\bibitem[{{Pereira} \& {Bryan}(2010)}]{Pereira2010}
{Pereira}, M.~J. \& {Bryan}, G.~L. 2010, \apj, 721, 939

\bibitem[{{Pereira} {et~al.}(2008){Pereira}, {Bryan}, \& {Gill}}]{Pereira2008}
{Pereira}, M.~J., {Bryan}, G.~L., \& {Gill}, S.~P.~D. 2008, \apj, 672, 825

\bibitem[{{Pereira} \& {Kuhn}(2005)}]{Pereira&Kuhn2005}
{Pereira}, M.~J. \& {Kuhn}, J.~R. 2005, \apjl, 627, L21

\bibitem[{{Piras} {et~al.}(2018){Piras}, {Joachimi}, {Sch{\"a}fer}, {Bonamigo},
  {Hilbert}, \& {van Uitert}}]{Piras2018}
{Piras}, D., {Joachimi}, B., {Sch{\"a}fer}, B.~M., {et~al.} 2018, \mnras, 474,
  1165

\bibitem[{{Robison} {et~al.}(2023){Robison}, {Hudson}, {Cuillandre}, {Erben},
  {Fabbro}, {Gavazzi}, {Guinot}, {Gwyn}, {Hildebrandt}, {Kilbinger},
  {McConnachie}, {Miller}, {Spitzer}, \& {van Waerbeke}}]{Robison2023}
{Robison}, B., {Hudson}, M.~J., {Cuillandre}, J.-C., {et~al.} 2023, \mnras,
  523, 1614

\bibitem[{{Rozo} {et~al.}(2016){Rozo}, {Rykoff}, {Abate}, {Bonnett}, {Crocce},
  {Davis}, {Hoyle}, {Leistedt}, {Peiris}, {Wechsler}, {Abbott}, {Abdalla},
  {Banerji}, {Bauer}, {Benoit-L{\'e}vy}, {Bernstein}, {Bertin}, {Brooks},
  {Buckley-Geer}, {Burke}, {Capozzi}, {Rosell}, {Carollo}, {Kind}, {Carretero},
  {Castander}, {Childress}, {Cunha}, {D'Andrea}, {Davis}, {DePoy}, {Desai},
  {Diehl}, {Dietrich}, {Doel}, {Eifler}, {Evrard}, {Neto}, {Flaugher},
  {Fosalba}, {Frieman}, {Gaztanaga}, {Gerdes}, {Glazebrook}, {Gruen},
  {Gruendl}, {Honscheid}, {James}, {Jarvis}, {Kim}, {Kuehn}, {Kuropatkin},
  {Lahav}, {Lidman}, {Lima}, {Maia}, {March}, {Martini}, {Melchior}, {Miller},
  {Miquel}, {Mohr}, {Nichol}, {Nord}, {O'Neill}, {Ogando}, {Plazas}, {Romer},
  {Roodman}, {Sako}, {Sanchez}, {Santiago}, {Schubnell}, {Sevilla-Noarbe},
  {Smith}, {Soares-Santos}, {Sobreira}, {Suchyta}, {Swanson}, {Thaler},
  {Thomas}, {Uddin}, {Vikram}, {Walker}, {Wester}, {Zhang}, \& {da
  Costa}}]{Rozo2016redMaGiC}
{Rozo}, E., {Rykoff}, E.~S., {Abate}, A., {et~al.} 2016, \mnras, 461, 1431

\bibitem[{{Schechter}(1976)}]{Schechter1976}
{Schechter}, P. 1976, \apj, 203, 297

\bibitem[{{Schneider} \& {Bridle}(2010)}]{SchneiderBridle2010}
{Schneider}, M.~D. \& {Bridle}, S. 2010, \mnras, 402, 2127

\bibitem[{{Schrabback} {et~al.}(2015){Schrabback}, {Hilbert}, {Hoekstra},
  {Simon}, {van Uitert}, {Erben}, {Heymans}, {Hildebrandt}, {Kitching},
  {Mellier}, {Miller}, {Van Waerbeke}, {Bett}, {Coupon}, {Fu}, {Hudson},
  {Joachimi}, {Kilbinger}, \& {Kuijken}}]{Schrabback2015}
{Schrabback}, T., {Hilbert}, S., {Hoekstra}, H., {et~al.} 2015, \mnras, 454,
  1432

\bibitem[{{Scoville} {et~al.}(2007){Scoville}, {Aussel}, {Brusa}, {Capak},
  {Carollo}, {Elvis}, {Giavalisco}, {Guzzo}, {Hasinger}, {Impey}, {Kneib},
  {LeFevre}, {Lilly}, {Mobasher}, {Renzini}, {Rich}, {Sanders}, {Schinnerer},
  {Schminovich}, {Shopbell}, {Taniguchi}, \& {Tyson}}]{Scoville2007}
{Scoville}, N., {Aussel}, H., {Brusa}, M., {et~al.} 2007, \apjs, 172, 1

\bibitem[{{Seljak}(2000)}]{Seljak2000}
{Seljak}, U. 2000, \mnras, 318, 203

\bibitem[{{Sheldon} {et~al.}(2004){Sheldon}, {Johnston}, {Frieman}, {Scranton},
  {McKay}, {Connolly}, {Budav{\'a}ri}, {Zehavi}, {Bahcall}, {Brinkmann}, \&
  {Fukugita}}]{Sheldon2004}
{Sheldon}, E.~S., {Johnston}, D.~E., {Frieman}, J.~A., {et~al.} 2004, \aj, 127,
  2544

\bibitem[{{Sif{\'o}n} {et~al.}(2015){Sif{\'o}n}, {Hoekstra}, {Cacciato},
  {Viola}, {K{\"o}hlinger}, {van der Burg}, {Sand}, \& {Graham}}]{Sifon2015}
{Sif{\'o}n}, C., {Hoekstra}, H., {Cacciato}, M., {et~al.} 2015, \aap, 575, A48

\bibitem[{{Singh} {et~al.}(2015){Singh}, {Mandelbaum}, \& {More}}]{Singh2015}
{Singh}, S., {Mandelbaum}, R., \& {More}, S. 2015, \mnras, 450, 2195

\bibitem[{{Singh} {et~al.}(2017){Singh}, {Mandelbaum}, {Seljak}, {Slosar}, \&
  {Vazquez Gonzalez}}]{Singh2017}
{Singh}, S., {Mandelbaum}, R., {Seljak}, U., {Slosar}, A., \& {Vazquez
  Gonzalez}, J. 2017, \mnras, 471, 3827

\bibitem[{Tinker {et~al.}(2010)Tinker, Robertson, Kravtsov, Klypin, Warren,
  Yepes, \& Gottloeber}]{Tinker2010}
Tinker, J.~L., Robertson, B.~E., Kravtsov, A.~V., {et~al.} 2010, The
  Astrophysical Journal, 724, 878

\bibitem[{{Troxel} \& {Ishak}(2015)}]{Troxel2015review}
{Troxel}, M.~A. \& {Ishak}, M. 2015, \physrep, 558, 1

\bibitem[{{Vakili} {et~al.}(2019){Vakili}, {Bilicki}, {Hoekstra}, {Chisari},
  {Brown}, {Georgiou}, {Kannawadi}, {Kuijken}, \& {Wright}}]{Vakili2019}
{Vakili}, M., {Bilicki}, M., {Hoekstra}, H., {et~al.} 2019, \mnras, 487, 3715

\bibitem[{{Vakili} {et~al.}(2023){Vakili}, {Hoekstra}, {Bilicki}, {Fortuna},
  {Kuijken}, {Wright}, {Asgari}, {Brown}, {Dombrovskij}, {Erben}, {Giblin},
  {Heymans}, {Hildebrandt}, {Johnston}, {Joudaki}, \& {Kannawadi}}]{Vakili2023}
{Vakili}, M., {Hoekstra}, H., {Bilicki}, M., {et~al.} 2023, \aap, 675, A202

\bibitem[{{van den Bosch} {et~al.}(2013){van den Bosch}, {More}, {Cacciato},
  {Mo}, \& {Yang}}]{vandenBosch2013-PaperI}
{van den Bosch}, F.~C., {More}, S., {Cacciato}, M., {Mo}, H., \& {Yang}, X.
  2013, \mnras, 430, 725

\bibitem[{{van Uitert} {et~al.}(2016){van Uitert}, {Cacciato}, {Hoekstra},
  {Brouwer}, {Sif{\'o}n}, {Viola}, {Baldry}, {Bland-Hawthorn}, {Brough},
  {Brown}, {Choi}, {Driver}, {Erben}, {Heymans}, {Hildebrandt}, {Joachimi},
  {Kuijken}, {Liske}, {Loveday}, {McFarland}, {Miller}, {Nakajima}, {Peacock},
  {Radovich}, {Robotham}, {Schneider}, {Sikkema}, {Taylor}, \& {Verdoes
  Kleijn}}]{vanUitert2016GAMAgroups}
{van Uitert}, E., {Cacciato}, M., {Hoekstra}, H., {et~al.} 2016, \mnras, 459,
  3251

\bibitem[{{van Uitert} {et~al.}(2015){van Uitert}, {Cacciato}, {Hoekstra}, \&
  {Herbonnet}}]{vanUitert2015}
{van Uitert}, E., {Cacciato}, M., {Hoekstra}, H., \& {Herbonnet}, R. 2015,
  \aap, 579, A26

\bibitem[{{van Uitert} {et~al.}(2017){van Uitert}, {Hoekstra}, {Joachimi},
  {Schneider}, {Bland-Hawthorn}, {Choi}, {Erben}, {Heymans}, {Hildebrandt},
  {Hopkins}, {Klaes}, {Kuijken}, {Nakajima}, {Napolitano}, {Schrabback},
  {Valentijn}, \& {Viola}}]{vanUitert2017b}
{van Uitert}, E., {Hoekstra}, H., {Joachimi}, B., {et~al.} 2017, \mnras, 467,
  4131

\bibitem[{{van Uitert} \& {Joachimi}(2017)}]{vanUitert2017}
{van Uitert}, E. \& {Joachimi}, B. 2017, \mnras, 468, 4502

\bibitem[{{Velliscig} {et~al.}(2015{\natexlab{a}}){Velliscig}, {Cacciato},
  {Schaye}, {Crain}, {Bower}, {van Daalen}, {Dalla Vecchia}, {Frenk},
  {Furlong}, {McCarthy}, {Schaller}, \& {Theuns}}]{Velliscig2015a}
{Velliscig}, M., {Cacciato}, M., {Schaye}, J., {et~al.} 2015{\natexlab{a}},
  \mnras, 453, 721

\bibitem[{{Velliscig} {et~al.}(2015{\natexlab{b}}){Velliscig}, {Cacciato},
  {Schaye}, {Hoekstra}, {Bower}, {Crain}, {van Daalen}, {Furlong}, {McCarthy},
  {Schaller}, \& {Theuns}}]{Velliscig2015}
{Velliscig}, M., {Cacciato}, M., {Schaye}, J., {et~al.} 2015{\natexlab{b}},
  \mnras, 454, 3328

\bibitem[{{Vlah} {et~al.}(2020){Vlah}, {Chisari}, \& {Schmidt}}]{Vlah2020}
{Vlah}, Z., {Chisari}, N.~E., \& {Schmidt}, F. 2020, \jcap, 2020, 025

\bibitem[{{Wright} {et~al.}(2019){Wright}, {Hildebrandt}, {Kuijken}, {Erben},
  {Blake}, {Buddelmeijer}, {Choi}, {Cross}, {de Jong}, {Edge},
  {Gonzalez-Fernandez}, {Gonz{\'a}lez Solares}, {Grado}, {Heymans}, {Irwin},
  {Kupcu Yoldas}, {Lewis}, {Mann}, {Napolitano}, {Radovich}, {Schneider},
  {Sif{\'o}n}, {Sutherland}, {Sutorius}, \& {Verdoes Kleijn}}]{Wright2019A}
{Wright}, A.~H., {Hildebrandt}, H., {Kuijken}, K., {et~al.} 2019, \aap, 632,
  A34

\bibitem[{{Wright} {et~al.}(2020){Wright}, {Hildebrandt}, {van den Busch},
  {Heymans}, {Joachimi}, {Kannawadi}, \& {Kuijken}}]{Wright2020}
{Wright}, A.~H., {Hildebrandt}, H., {van den Busch}, J.~L., {et~al.} 2020,
  \aap, 640, L14

\bibitem[{{Xia} {et~al.}(2017){Xia}, {Kang}, {Wang}, {Luo}, {Yang}, {Jing},
  {Wang}, \& {Mo}}]{Xia2017}
{Xia}, Q., {Kang}, X., {Wang}, P., {et~al.} 2017, \apj, 848, 22

\bibitem[{{Yang} {et~al.}(2003){Yang}, {Mo}, \& {van den Bosch}}]{Yang2003}
{Yang}, X., {Mo}, H.~J., \& {van den Bosch}, F.~C. 2003, \mnras, 339, 1057

\bibitem[{{Yang} {et~al.}(2007){Yang}, {Mo}, {van den Bosch}, {Pasquali}, {Li},
  \& {Barden}}]{Yang2007}
{Yang}, X., {Mo}, H.~J., {van den Bosch}, F.~C., {et~al.} 2007, \apj, 671, 153

\bibitem[{Zheng {et~al.}(2005)Zheng, Berlind, Weinberg, Benson, Baugh, Cole,
  Davé, Frenk, Katz, \& Lacey}]{Zheng2005}
Zheng, Z., Berlind, A.~A., Weinberg, D.~H., {et~al.} 2005, The Astrophysical
  Journal, 633

\bibitem[{{Zhou} {et~al.}(2023){Zhou}, {Tong}, {Troxel}, {Blazek}, {Lin},
  {Bacon}, {Bleem}, {Chang}, {Costanzi}, {DeRose}, {Dietrich}, {Drlica-Wagner},
  {Gruen}, {Gruendl}, {Hoyle}, {Jarvis}, {MacCrann}, {Mawdsley}, {McClintock},
  {Melchior}, {Prat}, {Pujol}, {Rozo}, {Rykoff}, {Samuroff}, {Sheldon}, {Shin},
  {Rosell}, {Yanny}, {S{\'a}nchez}, {Tucker}, {Sevilla-Noarbe}, {Zuntz},
  {Varga}, {Zhang}, {Alves}, {Amon}, {Bertin}, {Brooks}, {Burke}, {Kind}, {da
  Costa}, {Davis}, {De Vicente}, {Desai}, {Diehl}, {Doel}, {Everett},
  {Ferrero}, {Flaugher}, {Frieman}, {Gerdes}, {Gutierrez}, {Hinton},
  {Hollowood}, {Honscheid}, {James}, {Jeltema}, {Kuehn}, {Lahav}, {Lima},
  {Marshall}, {Mena-Fern{\'a}ndez}, {Menanteau}, {Miquel}, {Palmese},
  {Paz-Chinch{\'o}n}, {Pieres}, {Malag{\'o}n}, {Porredon}, {Raveri}, {Romer},
  {Sanchez}, {Smith}, {Soares-Santos}, {Suchyta}, {Swanson}, {Tarle}, {To},
  {Weaverdyck}, {Weller}, \& {Wiseman}}]{Zhou2023}
{Zhou}, C., {Tong}, A., {Troxel}, M.~A., {et~al.} 2023, \mnras, 526, 323

\end{thebibliography}


\appendix
\onecolumn
\section{Investigating how well  the model can predict the luminosity distribution of the galaxies} \label{A:L_distr}

\begin{figure}[!h]
    \centering
    \includegraphics[width=\textwidth]{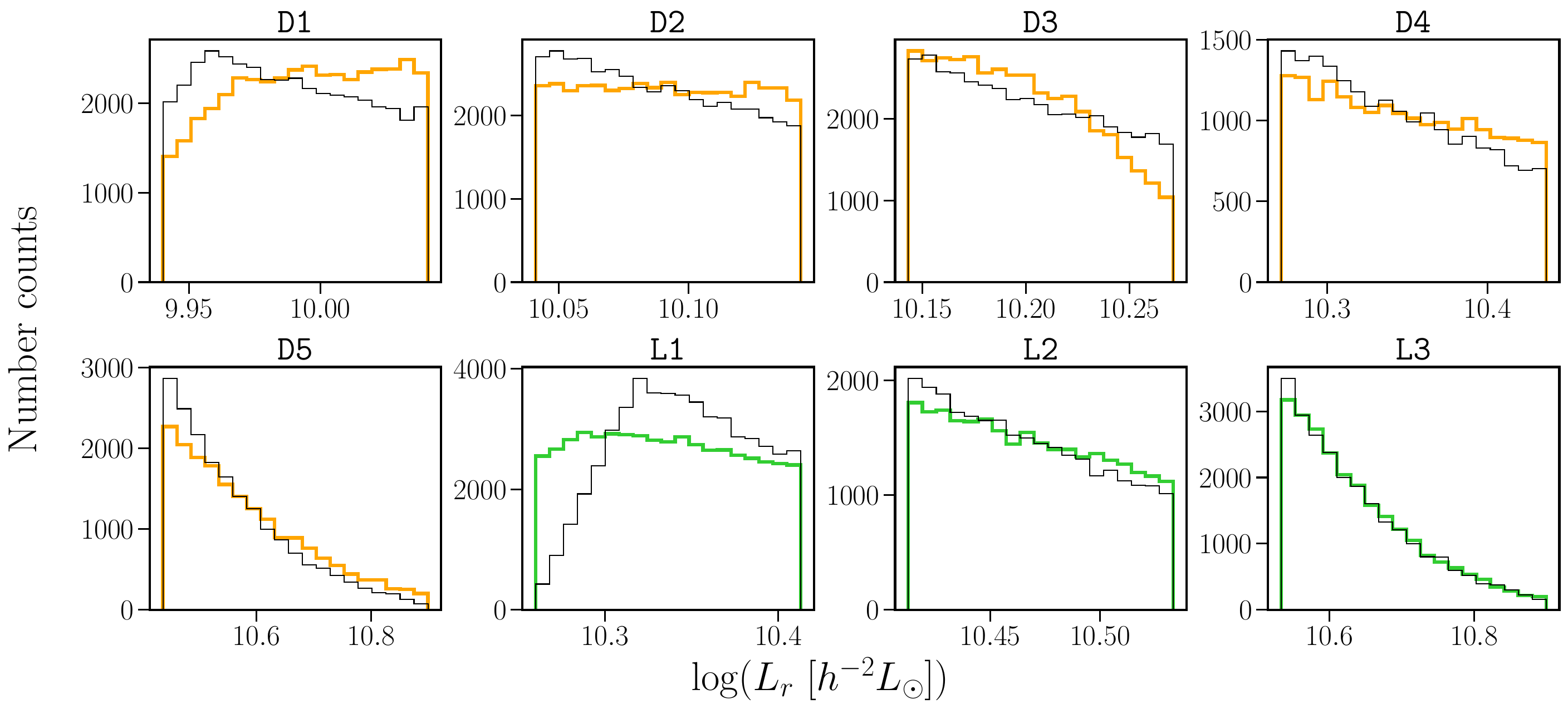}
    \caption{Real data number counts per luminosity bin (\dense: orange, \lum: green) and the prediction from the best-fit model (black), normalised for the number of objects in the given bin.}
    \label{fig:L_distr}
\end{figure}

The HOD modelling adopted in this work relies on the relation between the halo mass and the luminosity of the galaxies that populate it via the CLF. This allows us to test how well the best-fit model recovers the luminosity distributions of the galaxy samples, a quantity which is not directly used in the fit. This represents an independent test of the ability of the model to recover the properties of the lens galaxies. Figure~\ref{fig:L_distr} shows our results. We use the predicted luminosity function (LF) at the $z_{\rm eff}$ of the corresponding sample to generate the model distributions (black solid line), while the orange/green lines show the underlying true number counts.  Overall, the model reproduces sufficiently well the galaxy distribution per luminosity bin, with the exception of \texttt{L1}, where the distribution is significantly more peaked than the real one. 

\section{Fraction of physically associated galaxies} \label{A:fraction_of_associated_galaxies}

\begin{figure}[!h]
    \centering
    \includegraphics[width=0.45\columnwidth]{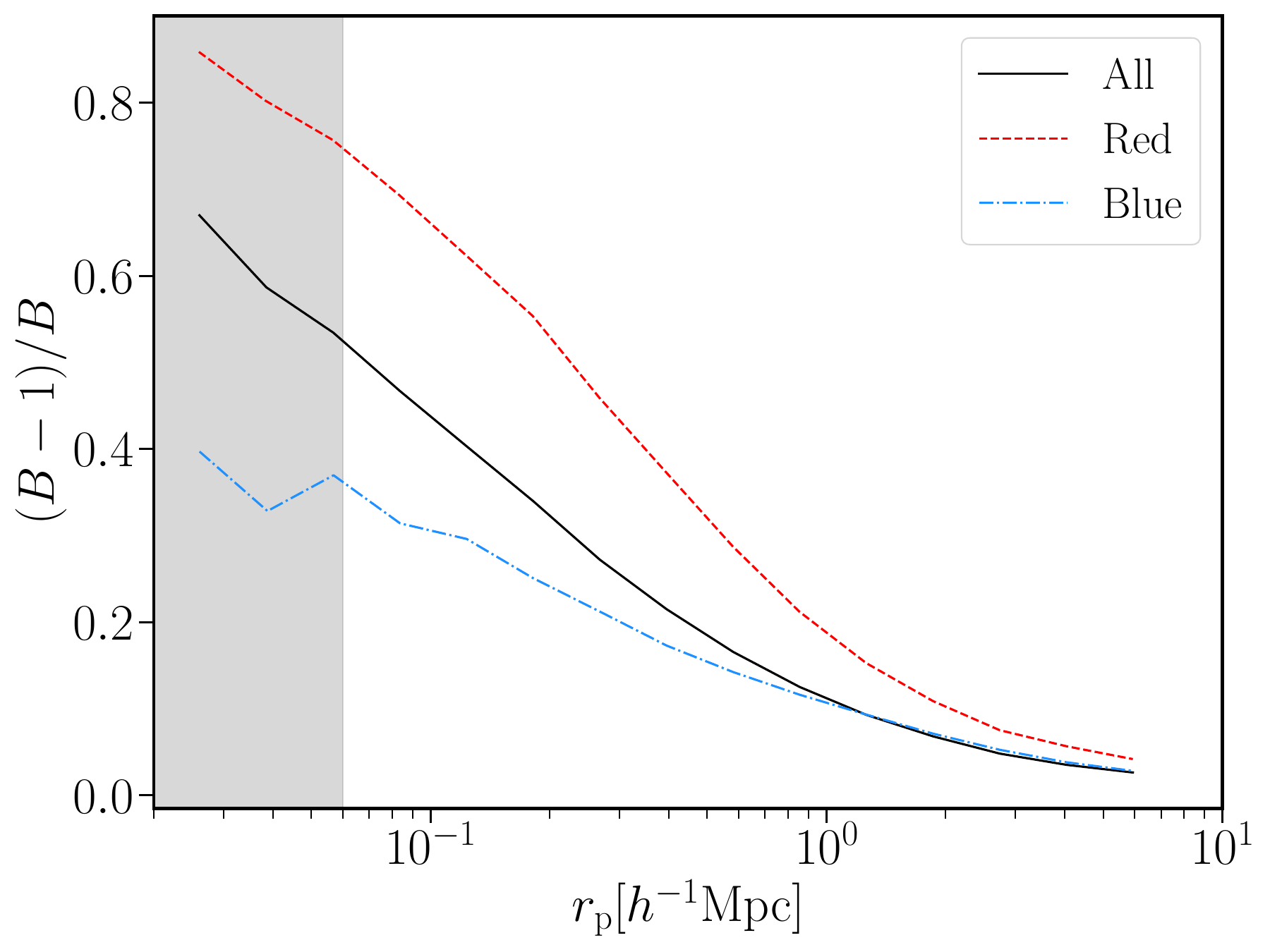}
    \caption{Fraction of source galaxies that are physically associated with the LRGs.}
    \label{fig:f_ass}
\end{figure}

In this appendix, we report the fraction of galaxies of the different source samples (all; red; blue) that are physically associated with the LRGs. The fractions are presented in Fig.~\ref{fig:f_ass} as a function of projected separation, and they are computed as $[B(\rp)-1]/B(\rp)$.

\end{document}